\documentstyle[preprint,aps]{revtex}
\begin{document}
\newcommand{\threej}[6]{
        \left(\! \begin{array}{ccc} {#1} & {#2} & {#3} \\ {#4} & {#5} & {#6} \end{array}\! \right) }
\newcommand{\sixj}[6]{ 
        \left\{ \begin{array}{ccc} {#1} & {#2} & {#3} \\ {#4} & {#5} & {#6} \end{array} \right\} }
\newcommand{\ninej}[9]{ \left\{ \begin{array}{ccc} {#1}&{#2}&{#3} \\ 
        {#4}&{#5}&{#6} \\ {#7}&{#8}&{#9} \end{array} \right\}}
\newcommand{\mx}[3]{ \langle\; {#1} \mid {#2} \mid {#3} \;\rangle }
\newcommand{\mxred}[3]{ \langle\; {#1} \|\; {#2} \;\| {#3} \;\rangle } 
\newcommand{\mxemp}[2]{ \langle\; {#1} \mid {#2} \;\rangle }
\newcommand{\iket}{ \mid l 1 1 M_i \;\rangle }
\newcommand{\mxba}[1]{ \mx{l' 1 1 M_f}{#1}{l 1 1 M_i} }
\newcommand{\mxbc}[1]{ \mx{l' 1 1 M_f}{#1}{L_C S_C J_C M_C} }
\newcommand{\mxca}[1]{ \mx{L_C S_C J_C M_C}{#1}{l 1 1 M_i} }
\newcommand{\mxredba}[1]{ \mxred{l' 1 1}{#1}{l 1 1} }
\newcommand{\mxredbc}[1]{ \mxred{l' 1 1}{#1}{L_C S_C J_C} }
\newcommand{\mxredca}[1]{ \mxred{L_C S_C J_C}{#1}{l 1 1} }
\newcommand{\bbra}{ \langle d_f \mid } 
\newcommand{\bra}[1]{ \langle~{#1}\!\mid } 
\newcommand{\ket}[1]{ \mid\! {#1}~\rangle }
\newcommand{\rbra}[1]{ \langle\; {#1} \| } 
\newcommand{\rket}[1]{ \| {#1}\; \rangle }
\newcommand{\aket}{ \mid d_i \;\rangle }
\newcommand{\fbra}{ \langle  l' 1 1 M_f \mid }
\newcommand{\matrixm}[1]{ {\mathcal{M}}_{fi}^{#1} }
\newcommand{\matrixt}[1]{ {\mathcal{T}}_{fi}^{#1} }
\newcommand{\eps}{ {\hat{\epsilon}}_{{\lambda}_i} }
\newcommand{\epspr}{ {\hat{\epsilon}}_{{\lambda}_f}^{\ast} }
\newcommand{\epsdp}{ \epspr\cdot\eps }
\newcommand{\kinit}{ {\vec{k}}_i }
\newcommand{\kfin}{ {\vec{k}}_f }
\newcommand{\khatinit}{ {\hat{k}}_i }
\newcommand{\khatfin}{ {\hat{k}}_f }
\newcommand{\vsh}[3]{ {\vec{T}}_{ {#1}\,{#2}\,{#3} }(\hat{r})}
\newcommand{\winit}{ {\omega}_i }
\newcommand{\muinit}{ {\lambda}_i }
\newcommand{\wfin}{ {\omega}_f }
\newcommand{\mufin}{ {\lambda}_f }
\newcommand{\epsexp}{ {\hat{\epsilon}}_{\lambda} e^{i\vec{k}\cdot\vec{r}} }
\newcommand{\sss}{\scriptscriptstyle}
\newcommand{\wignerd}{D_{M\lambda}^{(L)}(0,-\vartheta,-\varphi) }
\newcommand{\wignerdneg}{D_{M-\lambda}^{(L)}(0,-\vartheta,-\varphi) }
\newcommand{\wignerdi}{D_{M\,\muinit}^{(L)}(0,0,0) }
\newcommand{\wignerdf}{D_{M'\,-\mufin}^{(L')}(0,-\theta,-\phi) }
\newcommand{\wignerdfnom}{D_{M_f-M_i-\muinit,-\mufin}^{(L')}(0,-\theta,-\phi) }
\newcommand{\green}{G_{\sss L_C}(r,r';E_0) }
\newcommand{\greenpr}{G_{\sss L_C}(r,r';E_0') }
\newcommand{\rmass}{\mu_{\sss {\mathrm red} } }
\newcommand{\dum}{\xi}
\newcommand{\iso}[1]{ \tilde{#1} }
\newcommand{\agen}{ \bar{\alpha} }
\newcommand{\bgen}{ \bar{\beta} }
\newcommand{\imag}{ \mathrm{Im} }

\title{
\begin{flushright}
{ \normalsize nt@uw-99-6
 Jan.  1999}
\end{flushright}
   Neutron Polarizabilities From 
Compton Scattering on the   Deuteron?          }
\author{Jonathan  J. Karakowski and Gerald  A. Miller  }
\address{Department of Physics, Box 351560\\
University of Washington \\
Seattle, WA 98195-1560, U.S.A.}

\maketitle
\begin{abstract}A calculation of deuteron Compton scattering using
  non-relativistic
perturbation theory is presented,
with the primary motivation of investigating the feasibility
of determining the neutron polarizabilities from this type of experiment.  
This calculation is expected to be valid for energies below 100 MeV.
 Pion-exchange,  relativistic, and recoil corrections are also included.
 The low-energy theorem
for gauge invariance is shown to be satisfied.   The relative
effects of the different terms  and their effects on 
the determinations of the polarizabilities
are discussed at energies of 49, 69, and 95 MeV.    
The cross-section is dominated by the seagull, polarizability,
and electromagnetic multipole
interactions.  Relativistic and pion-exchange
terms are also important, while recoil corrections
and multipoles of $L$=2 and greater are negligible.
The calculation provides a reasonable description
of the experimental data points at 49 and 69 MeV.
The polarizabilities are difficult to determine at these energies. 
A more accurate determination of the polarizabilities
may be  possible  at  95 MeV. 
\end{abstract}

\section{Introduction}
This paper is concerned with determining the polarizability of the neutron
using
Compton scattering
from a deuteron.
 This is not a new idea and there have been previous calculations.
However,
advances
on both the theoretical and experimental fronts invite the possibility 
of not only a more accurate calculation, but a more accurate measurement.   
The intent of this work is provide a complete potential model  calculation.

A deuteron
Compton scattering
experiment was performed in 1994 \cite{lu94}, using 49 and 69 MeV photons, 
and experiments are in progress at
Saskatoon \cite{ho98} and Lund \cite{lund98}.
This has spurred
 recent activity.   Brief
articles by
Wilbois, Wilhelm, and Arenh{\"{o}}vel,\cite{ar95}
and L'vov and
Levchuk \cite{lv95,lv98} quote new theoretical results for Compton scattering.
Calculations
using effective field theory,
based on the approach of Kaplan, Savage and Wise\cite{ksw},
have also been performed \cite{ch98,chen98}.
    Another calculation\cite{bmpv99},
using
an approach intermediate between ours and that of Ref.~\cite{ch98,chen98},
is nearly complete.    

Here is an outline of  this paper. The remainder of this
introduction reviews the concept of 
 polarizability and examines the 
previous theoretical and experimental work.  
Descriptions of the methods used in the calculation are discussed in Sect. II.
Checks which were performed to ensure the correctness of the work are
also detailed there.  The  results of the numerical calculation are
presented in Sect.~III. The possibility of
determining 
the  neutron polarizabilities   from deuteron Compton
scattering is detailed.  The conclusion and summary are  in Sect.~IV.
  Comparisons between our  and previous    
calculations will be made in the summary. Some of the 
details of calculations omitted in the main text  are in the Appendices. 
Complete  formulae for all of the relevant scattering amplitudes may be found
in the 1998 Ph.~D. thesis\cite{thesis} 
of one of the authors. This work is based on that thesis.

\subsection{Background and Motivation \label{ch:int} }
The symbol $\alpha$ is traditionally used to represent the
electric polarizability, and $\beta$ is used for the 
magnetic polarizability.
The polarizability is the constant of
proportionality between an external field and the average dipole
moment that it induces.
  This induced moment
in turn changes the potential energy of a system by an amount
$
\Delta E = -\frac{1}{2} \alpha E^2 - \frac{1}{2} \beta B^2, \label{eq:c1-1}
$
known as the quadratic Stark effect.
The  nucleon
polarizabilities are  obtained using 
 Hamiltonians for a dipole interacting
with an external field: 
$
H_{\mathrm{int}} = -d_z E,
$ or $-\mu_z B.$
Second order perturbation theory leads to
\begin{equation}
\alpha = 2 \sum_{N'} \frac{ \left| \mx{N}{d_z}{N'} \right|^2 }{E_{N'}-E_N},
\quad
\beta = 2 \sum_{N'} \frac{ \left| \mx{N}{\mu_z}{N'} \right|^2 }{E_{N'}-E_N}.
\label{eq:c1-2}
\end{equation}
We  make an   order-of-magnitude estimate of
$\alpha$ by letting $N'$ be an $N^{\ast}$ state of mass 1440 MeV and taking
the square of the matrix element to be
about $\frac{1}{3} \mathrm{fm}^2$. Then  
$\alpha \approx 13 \times 10^{-4} \mathrm{fm}^3.$
 This is of the same order as the currently accepted value of
$\alpha \approx 12 \times 10^{-4} \mathrm{fm}^3$.
Hereafter the unit of polarizability will be $10^{-4} \mathrm{fm}^3$ .

 
The above
expressions define the polarizability in terms of either  an energy shift
or an induced dipole moment.
A third definition, which turns out to be the most practical,
involves the scattering amplitude for elastic photon scattering,
commonly known as Compton scattering.
An initial   photon, of  laboratory four-momentum
$(\omega_i , \vec{k}_i)$
and polarization vector $\eps$, where $\lambda_i = \pm 1$, is absorbed by
a nucleon, which emits a final photon of four-momentum 
$(\omega_f , \vec{k}_f)$ and
$\hat{\epsilon}_{\lambda_f}$.  
The angle between $\kinit$ and $\kfin$ is labeled $\theta$,
and the initial and final photon 
energies are related by 
$ 
\wfin =  \winit /( 1 + \frac{\winit}{m_N}(1 - \cos{\theta}) ).
$                               
The scattering amplitude
(to order $\omega^2$) is given  as  \cite{pe81}
\begin{eqnarray}
f(\omega, \winit) & = & -\frac{e^2}{m_N}
(\epsdp )
+ (\winit + \wfin) g(e,m_N,\kappa_N; \hat{k}_f,\epspr,\hat{k}_i,\eps)
+ \nonumber\\
& & \ \ \ \ \ \ \ \     \bar{\alpha} \wfin\winit (\epsdp ) + \bar{\beta}
\wfin\winit (\hat{k}_f \times \epspr) \cdot
        (\hat{k}_i \times \eps) +  \nonumber\\
& &
\ \ \ \ \ \ \ \ \ \ \ \ \ \ \ \ \wfin\winit
h(e,m_N,\kappa_N; \hat{k}_f,\epspr,\hat{k}_i,\eps). \label{eq:c1-3}
\end{eqnarray}
\noindent The first term, the Thomson amplitude, 
is the dominant one
at low energies.  
The 
internal structure becomes important only in second order. 
The terms with the angular dependence $ (\epsdp )$ and
$ (\hat{k}_f \times \epspr) \cdot (\hat{k}_i \times \eps)$
are written separately, and all other combinations
of the four photon unit vectors are lumped together in the function $h$.
The coefficients 
of these two terms are defined as  the
generalized polarizabilities $ \bar{\alpha} $ and $  \bar{\beta} $.

The generalized polarizabilities 
 $ \bar{\alpha} $, $  \bar{\beta} $ are not the
 same as the static polarizabilities
$\alpha$,$\beta$.  They are related by
$
\bar{\alpha} = \alpha + \Delta\alpha,\quad
\bar{\beta} =\beta+\Delta\beta $, 
\noindent where $\Delta\alpha,\Delta\beta$ arise from
terms in the scattering amplitude of
the same energy and angular dependence as those of $\alpha,\beta$.
The first term in the explicit  
expression for $\Delta\alpha$ is \cite{pe63,ma90} :
$ 
\Delta\alpha = \frac{e^2r_e^2}{3m} + \frac{e^2(1+\kappa^2)}{4m^3},
$
where $r_e$ stands for the charge radius, and 
accounts for about 40\% of $\agen_p$. This 
 is only about a 5\% effect in the
neutron.  
The first term in $\Delta\beta$ is \cite{ma90}
$
\Delta\beta = -\frac{e^2 e_e^2}{6m}.
$ 
Additional corrections involve
the nucleon form factors $F_1(4m^2), F_2(4m^2)$ and their derivatives
\cite{ma90}.

Our primary  interest is  in determining polarizabilities through
experiment, but it is useful
to review  the  theoretical calculations.
Baldin \cite{ba60} obtained
an estimate $
4.0 \le \agen_p \le 15.0,
$ for the proton polarizabilities by
including the first excited state ($N' = N + \pi$).
Gell-Mann and Goldberger \cite{ge54} derived
the once-subtracted dispersion relation for
a forward scattering amplitude $f(\omega)$.
The low energy limit of   Eq.~(\ref{eq:c1-3}),
 along with the  dispersion relation   and the
the optical theorem, leads to the Baldin sum rule \cite{ba60}:
\begin{equation}
\agen + \bgen = \int_{m_{ \pi} }^{\infty}
\frac{ \sigma_{\mathrm{tot}}(\omega' ) d\omega' }{2\pi^2 \omega^{\prime 2}},
\end{equation} 
where $\sigma_{\mathrm{tot}}(\omega )$ is the total
cross-section for photoabsorption.
This relation provides a model-independent
constraint on
the possible values of $\agen$ and $\bgen$.
 The generally accepted results from the sum rule are \cite{da70,sc80}
$
\agen_p + \bgen_p  =  14.3 \pm 0.5, 
\agen_n + \bgen_n  =  15.8 \pm 0.5. 
$         However, a recent experiment \cite{ba98} predicts:
$
\agen_p + \bgen_p  = 13.69 \pm 0.14, 
\agen_n + \bgen_n  =  14.44 \pm 0.69.
$
The values for the proton are consistent with the older ones,
but the new neutron values are 
slightly lower.

A dispersion relation for
$\agen - \bgen$ can also be derived \cite{be74,be77},
but it contains a contribution that depends on the amplitudes for
$\gamma\gamma \rightarrow \pi\pi$, which are not well known.
Nevertheless, this
difference has been estimated to be \cite{ho94}
$
\agen_p - \bgen_p  \approx 3.2, 
\agen_n - \bgen_n  \approx  3.9. 
$
Comparison with the sum rules above yield smaller values for the
individual polarizabilities than
the $\agen_p \approx 11$ and $\agen_n \approx 12$ measured 
experimentally.

Various quark models have also been used.  
A simple nonrelativistic model with a harmonic oscillator potential yields
reasonable values for $\bgen$ but predicts
$\agen_p > \agen_n$,  so more sophisticated models are needed.
For example, Werner and Weise \cite{we85} obtained $\agen$ and $\bgen$
using a valence quark core surrounded by a pion cloud to be
$
\agen_N  \approx  7-9,
\bgen_N  \approx 2.$
Perhaps the most promising new method is to  use  chiral perturbation theory.
Calculations with only the one-loop contribution produce \cite{me92}:  
$
\agen_N = 10 \bgen_N \approx 13.6$.
 Including the next order as well as the effects of the $\Delta$
resonance gives \cite{me93,me94}
$\agen_p = 10.5 \pm 2.0,\quad$  $\agen_n = 13.4 \pm 1.5,\quad$ 
$\bgen_p = 3.5 \pm 3.6,\quad $  $\bgen_n = 7.8 \pm 3.6.$ 
If we compare these two calculations with the Baldin sum rule, we find
agreement
for the predictions for the proton.  However, the earlier calculation 
gives better agreement for the neutron than the higher-order one.
The most important 
contributions to $\agen$ come from polarizing the
pion cloud - about 50-70\% \cite{we85}.
The small value of $\bgen \approx 2$ can
be attributed to cancellations between the large
paramagnetic contribution from the $\Delta$ resonance ($\beta$) 
and the diamagnetic contribution from the
pion cloud ($\Delta\beta$) \cite{mu93}.

 Several experiments have been performed
over the past few years with the aim of measuring nucleon polarizabilities.
Most of these
have specifically targeted the proton, with a
differential cross-section for Compton scattering
given  (to order $\omega^2$) as
\begin{equation}
\frac{d\sigma}{d\Omega}  =  \frac{d\sigma}{d\Omega}_{\mathrm{Born}} 
 - \frac{\alpha}{m_p} \left( \frac{\omega '}{\omega} \right)^2 \omega\omega ' 
\left\{ \frac{1}{2} (\agen_p + \bgen_p ) (1 + \cos\theta)^2 + \frac{1}{2}
  (\agen_p - \bgen_p ) (1 -\cos\theta)^2 \right\}, \label{eq:c3-2}
\end{equation}
\noindent where $\alpha$ is the fine structure constant, and the Born term
is that of Powell\cite{po49}.
Note that the polarizability terms in equation~(\ref{eq:c3-2}) arise from
interference between the polarizability
amplitude of equation~(\ref{eq:c1-3}) and the
Thomson amplitude.
This formula shows  the sensitivity of the cross-section to $\agen_p$ and 
$\bgen_p$ at different angles.
The cross-section at forward angles is most sensitive to $\agen_p+\bgen_p$,
while at backward angles only $\agen_p-\bgen_p$ can be measured.
At $90^{\circ} $ the $\beta_p$ terms drop out
completely. 
The expansion to order $\omega^2$ is 
only valid to about 100 MeV \cite{na94}.
Since the polarizability terms are of this order, 
too low an energy means
that the cross-section is does not depend on  $\agen_p$ and $\bgen_p$.  
Thus the  optimal energy range for an 
experiment
would be 70-100 MeV, which  balances sensitivity with expansion validity.  

A summary of  a few  proton Compton scattering experiments
is given in Table~\ref{t:c3-1} below. The data are those of 
 Moscow experiment \cite{ba75}, 
Illinois group \cite{fe91},\cite{ma95}, and
Mainz experiment \cite{ze92}. 
More extensive reviews of these
and other proton Compton scattering experiments
can be found in the literature \cite{pe81,na94}.
Despite
some shortcomings, measurements for the polarizabilities in
these experiments are in reasonable agreement with each other and 
with theoretical predictions.  This is not the case for the neutron. 

A direct  neutron Compton scattering experiment would require
the use of a neutron target
to detect a very small cross section.               
Therefore Coulomb scattering and quasi-elastic Compton scattering
from a neutron bound in the deuteron have been used.
The large Coulomb field of  a heavy nucleus can induce a  
large dipole
moment. 
Two separate 1988 experiments, both using
$\left.\right.^{208}$Pb as the target nucleus,  
quote large uncertainties:
$
\alpha_n  =  12.0 \pm 10.0, $  \cite{sh88} 
$\alpha_n  =  8.0 \pm 10.0$  \cite{ko88} 
A more recent experiment by Schmiedmayer et al. \cite{sh91},
also using a lead target, produced
$
\alpha_n = 12.0 \pm 1.5 \pm 2.0.
$
This is the value normally quoted for the neutron electric polarizability.  
However,
the accuracy of this result has been questioned \cite{en97,ni92}.
It has been claimed that the uncertainties
were underestimated,  and that the best estimate for $\alpha_n$ is
really $\sim 7-19$ \cite{en97}.  
An experiment to measure quasi-free Compton scattering
by the neutron bound in the deuteron\cite{lv}
 was carried out by
Rose et al. \cite{ro90} in 1990, and obtained 
$
\alpha_n = 11.7^{+4.3}_{-11.7}.
$
The uncertainties are large, but  recent  
arguments indicate
that this method  is capable of producing 
more accurate results \cite{wi98} and should be revisited.

It is clear that better experimental measurements of the 
neutron polarizability are needed.  
This is why we have chosen to investigate deuteron Compton
scattering.
This idea seems to have been
originally
suggested by Baldin \cite{ba60}, who calculated the
cross-section in the impulse 
approximation.   A more extensive calculation was undertaken by Weyrauch
\cite{we88,we90},
but certain deficiencies  to be discussed below,
as well 
as a lack of emphasis on the neutron polarizability, suggest that this question
can be reexamined.   
\section{Deuteron Compton Scattering Calculations \label{ch:calc} }

The Feynman diagrams for several of the processes contributing to deuteron 
Compton scattering are shown in  Fig.~\ref{fig:d1-1}.  The
ovals at either end
represent the deuteron wavefunctions, while the solid lines are the  
individual nucleons with which the photons (wavy lines) interact.
Not all possible combinations
are shown; interactions can occur on either nucleon with either an
incoming or outgoing photon.
 Figure~\ref{fig:d1-1}(a) is known as the seagull diagram,
arising
from the term in the Hamiltonian that is proportional to $A^2$.
All other one-body interactions,
which are at least of order  $\omega^1$, are represented by
Fig.~\ref{fig:d1-1}(b).
These include the terms which depend on the polarizabilities
$\alpha$ and $\beta$, as well 
as some relativistic corrections.
Dispersive diagrams
without meson exchange are depicted in  Fig.~\ref{fig:d1-1}(c)-(d).
At lowest order,
each vertex
can be either an electric or magnetic multipole  interaction.
Our notation is that the intermediate
state includes the effects of the un-denoted
n-p interaction. Two
of the meson-exchange effects which are included in this calculation
are shown in  Figure~\ref{fig:d1-1}(e)-(f).
The term of Figure~\ref{fig:d1-1}(f)
is called the ``vertex correction''.

\subsection{Overview of Calculation}

We  begin with an overview of the methods used in the calculation.  
The differential cross-section will be calculated non-relativistically, to
second order in the electric charge,
using Fermi's golden rule.
The transition matrix $ {\mathcal T}_{fi}$ is then given by 
\begin{eqnarray}
  {\mathcal T}_{fi}  =
  \langle d_f, \vec{P}_f, \gamma_f\mid{H^{\mathrm{int}} }\mid d_i, 
\vec{P}_i, \gamma_i\rangle + 
& & \sum_{C, \vec{P}_C} \frac{ \mx{d_f, \vec{P}_f, \gamma_f}
  {H^{\mathrm{int}}}{C, \vec{P}_C}
\mx{C, \vec{P}_C}{H^{\mathrm{int}}}{d_i, \vec{P}_i, \gamma_i} }{E_{d_i} + 
P_i^2/2m_d + \hbar\winit - E_C - P_C^2/2m_d + i\varepsilon} + \cdots,
\label{eq:d2-2}
\end{eqnarray}
where the $\cdots$ represents the crossing term, see
Eq.~(\ref{eq:mainterm1}). 
The notation $d_{i(f)}$ is meant to represent all
quantum numbers needed to describe the
initial (final) deuteron state, except for the
center-of-mass momentum $\vec{P}$, which will be
absorbed into a momentum-conserving 
delta function.  $C$ represents all possible
intermediate  states.

The most important terms we include are represented by 
a non-relativistic
Hamiltonian:
\begin{eqnarray} 
\lefteqn{H^{\mathrm{int}}
  =  \sum_{j=n,p} \left[ \frac{e_j^2}{2m_j} A^2(\vec{x}_j) -
\frac{e_j}{m_j} \vec{A}(\vec{x}_j)\cdot\vec{p}_j -  \right.
 \frac{e(1+\kappa_j)}{2m_j} \vec{\sigma}_j\cdot
 \left( \vec{\nabla}_j \times\vec{A}(\vec{x}_j) \right)- }\label{eq:d2-3} \\      & &  \frac{1}{2} \bar{\alpha}_j \left( \frac{ \partial\vec{A}(\vec{x}_j) }{\partial t } \right)^2 
-  
\frac{1}{2} \bar{\beta}_j  \left( \vec{\nabla}_j \times\vec{A}(\vec{x}_j)\right)^2
+  \frac{if_{\pi}e_{\pi}}{m_{\pi}} \left( \vec{\sigma}_j \cdot
\vec{A}(\vec{x}_j) \right) \left( \iso{\tau}_j \cdot \iso{\phi}(\vec{x}_j) \right) + \nonumber\\
& & \left.\frac{f_{\pi}\hbar}{m_{\pi}} \left( \vec{\sigma}_j \cdot
\vec{\nabla}_j \right) \left( \iso{\tau}_j \cdot \iso{\phi}(\vec{x}_j) \right) \right]
+ \frac{1}{\hbar^2}\int d^3xe^2 A^2(\vec{x}) \left[ \phi_+(\vec{x})
\phi_-(\vec{x}) + \phi_-(\vec{x})\phi_+(\vec{x}) \right], \label{apph}
\nonumber
\end{eqnarray} 
where the vector potential is given by
\begin{equation}
 \vec{A}(\vec{x_j}) = \frac{1}{\sqrt{V} } \sum_{\vec{k},\lambda = \pm 1 } \sqrt{\frac{2\pi\hbar }{\omega } } [ a_k
{\hat{\epsilon}}_{\lambda} e^{i\vec{k}
\cdot{\vec{x}}_j } + a_k^{\dagger} {\hat{\epsilon}}_{\lambda}^{\ast} e^{-i\vec{k}\cdot{\vec{x}}_j } ] .
\end{equation}
The operators  $a_{k,\lambda}^{(\dagger )}$ destroy (create) a photon with momentum $\vec{k}$
and polarization $\lambda $.
 Similarly, the pion field operator is given by 
\begin{equation}
\phi_{\pm}(\vec{x}_j) = \sum_q \frac{\hbar}{\sqrt{V}} \sqrt{ \frac{2\pi}{E_{\pi}} }
\left( a_{\mp} e^{i\vec{q}\cdot\vec{x}_j} + a_{\pm}^{\dagger} e^{-i\vec{q}\cdot\vec{x}_j} \right).
\end{equation}
Here, $a_{\pm}$ destroys a $\pi_{\pm}$.  The tildes in the Hamiltonian
indicate vectors in isospin space.  The approximate Hamiltonian (\ref{apph})
includes the effects of electric and magnetic dipole interactions. As
discussed in Sect.~IIC
below and in Ref.~\cite{thesis},  higher multipole moments are included also. 

The transition matrix is proportional to  a momentum-conserving 
delta function.  We remove this delta function and   define  the 
scattering amplitude $\matrixm{}$: 
\begin{equation}
 \matrixt{} = \ \frac{1}{V} \frac{2\pi\hbar}{\sqrt{\winit\wfin}}
 \label{eq:d3-6}
\frac{ {\delta}({\vec{P}}_f + \kfin - {\vec{P}}_i - \kinit)}{V} \matrixm{} . 
\end{equation}   
The differential cross-section is given by
\begin{eqnarray}
\left(\frac{d\sigma}{d\Omega}\right)_{fi} & = &
\left( \frac{\wfin}{\winit} \right)^2 \frac{E_f}{m_d} \mid\matrixm{} \mid^2 , 
\end{eqnarray}
where $E_f =\winit + m_d - \wfin $, $m_d$ is the deuteron mass.
The spin-averaged cross-section is computed by summing 
over the final and averaging over the initial polarizations.

\subsection{Seagull Terms}

 Using the
interaction Hamiltonian
$H^{SG} = \frac{e^2}{2 m_p} A^2(\vec{x_p}) 
$ accounts for the seagull diagram.  The resulting contribution to the
  transition matrix is given by
\begin{eqnarray}
 \matrixm{SG} & = & \frac{\sqrt{12\pi}e^2}{m_p} (\epsdp) (-1)^{1-M_f} \times\label{eq:d3-4}\\
& & \ \ \ \ \ \ \left[ \threej{1}{0}{1}{-M_f}{0}{M_f} Y_{00}^{\ast}(\hat{q}) \right.  (I_0^{00}+I_0^{22}) {\delta}_{M_f, M_i} +
\nonumber \\
& & \ \ \ \ \ \ \ \ \left. \threej{1}{2}{1}{-M_f}{M_f-M_i}{M_i} Y_{2, M_f-M_i}^{\ast}(\hat{q}) (I_2^{02} +
I_2^{20} - \frac{I_2^{22}}{\sqrt{2}}) \right], \nonumber
\end{eqnarray}
where $\vec{r}  \equiv \vec{x}_p - \vec{x}_n, $ and
\begin{equation}
 I_L^{l'l} \equiv \int_0^{\infty} dr u_{\sss l'}(r)
 j_{\sss L}(\frac{qr}{2}) u_{\sss l}(r).
\label{sgt}\end{equation}
The radial deuteron wavefunction is represented by $u_{0,2}$ 
 in which
 $0,2$ refer to the orbital angular momentum.  We use
the ``B'' wavefunctions of Ref.~\cite{ma89},
but the numerical results do not depend on
this choice.

The terms that depend on the nucleon polarizabilities are very similar to the
seagull term.  Their transition matrices are given by the following
expressions:
\begin{eqnarray}
\matrixm{\alpha} & = &
-\wfin\winit\sqrt{12\pi} (\epsdp) (-1)^{1-M_f} ( {\alpha}_p + {\alpha}_n )  
\label{eq:alpha2}\times \\
 & & \ \ \ \left[ \threej{1}{0}{1}{-M_f}{0}{M_f} Y_{00}^{\ast}(\hat{q})
   (I_0^{00} + I_0^{22})\delta_{M_f,M_i} 
\right. + \nonumber\\  
 & & \ \ \ \ \ \ \left. \threej{1}{2}{1}{-M_f}{M_f-M_i}{M_i}
   Y_{2,M_f-M_i}^{\ast}(\hat{q})(
I_2^{02} + I_2^{20} - \frac{I_2^{22}}{\sqrt{2}} ) \right], \nonumber
\end{eqnarray}
\begin{eqnarray}
\matrixm{\beta} & = & -\wfin\winit \sqrt{12\pi}
\left[ \left( \khatinit\times\eps \right) \cdot \left(\khatfin 
\times\epspr\right) \right] (-1)^{1-M_f}
({\beta}_p+{\beta}_n)\times\nonumber \\
& &\ \ \ \left[ \threej{1}{0}{1}{-M_f}{0}{M_f}
  Y_{00}^{\ast}(\hat{q})(I_0^{00} + I_0^{22}) 
\delta_{M_f,M_i} + \right. \nonumber\\
& &\ \ \ \ \ \ \left.
  \threej{1}{2}{1}{-M_f}{M_f-M_i}{M_i} Y_{2,M_f-M_i}^{\ast}(\hat{q}) 
(I_2^{02} + I_2^{20} - \frac{I_2^{22}}{\sqrt{2}}) \right].
\end{eqnarray}

\subsection{Dispersive Terms}

Most of the remaining terms to be calculated are the
dispersive terms, which are
second-order in the interaction Hamiltonian.
The contribution of the dispersive term to the transition matrix
has the form
\begin{eqnarray}{\cal T}_{fi}^{disp}
  & = & \sum_{C, \vec{P}_C} \left\{ \frac{ \mx{d_f, \vec{P}_f ,{\gamma}_f}{H^{\mathrm{int} }}{C, \vec{P}_C}
\mx{C, \vec{P}_C}{H^{\mathrm{int}}}{d_i, \vec{P}_i, {\gamma}_i} }{\hbar{\omega}_i+E_{d_i}-E_C-
P_C^2/2m_d +i\varepsilon}
+ \right.  \label{eq:d4-1}\\
& & \left. \frac{ \mx{d_f, \vec{P}_f, {\gamma}_f}{H^{\mathrm{int}}}{C, \vec{P}_C, {\gamma}_i, {\gamma}_f }
\mx{C, \vec{P}_C, {\gamma}_i, {\gamma}_f }{H^{\mathrm{int}}}{d_i, \vec{P}_i, {\gamma}_i} }
{-\hbar{\omega}_f+E_{d_i}-E_C-P_C^2/2m_d+i\varepsilon} \right\} \nonumber ,
\end{eqnarray}
in which 
$H^{\mathrm{int}}
= -\int \vec{J}(\vec{\dum})\cdot \vec{A}(\vec{\dum}) d^3\dum$.
We use
$\vec{\dum}$  as a dummy variable, while $\vec{x}$ is a nucleon variable
and $\vec{r}$ is a deuteron variable.
The term  $\vec{A}$ is handled using a multipole expansion:
\begin{eqnarray}
\lefteqn{ \epsilon_{\lambda} e^{i\vec{k}\cdot\vec{\dum} } =
 \sum_{L = 1}^{\infty} \sum_{M=-L}^{L}\wignerd i^L\sqrt{\frac{2\pi(2L+1)}{L(L+1)}}\times} \label{eq:d4-2}\label{multipole}\\
& &  \left\{ -\frac{i}{\omega}\vec{\nabla_{\dum}} \left( 1+\dum\frac{d}{d\dum} \right) j_{\sss L}(\omega\dum) 
Y_{LM}(\hat{\dum}) -
i\omega \vec{\dum} j_{\sss L}(\omega\dum) Y_{LM}(\hat{\dum}) -
\lambda \vec{L}  Y_{LM}(\hat{\dum})  j_{\sss L}(\omega\dum) \right\} ,
\nonumber
\end{eqnarray}
where
$\wignerd$ is the Wigner-d function (we
use the convention of \cite{rose57}) ,
which is the overlap of the state $\ket{L\lambda}$, rotated by
the Euler angles $(0,-\vartheta,-\varphi)$, with the state $\ket{LM}$.  
To simplify the notation, we define  functions $\Phi_i$ and $\Phi_f$ by
\begin{eqnarray}
\frac{1}{\sqrt{V}} \sqrt{\frac{2\pi\hbar}{\winit} } \eps
e^{i\kinit\cdot\vec{\dum}} & = & 
\vec{\nabla}\Phi_i(\vec{\dum}) + \cdots, \label{eq:d4-80} \\
\frac{1}{\sqrt{V}} \sqrt{\frac{2\pi\hbar}{\wfin}} \epspr
e^{-i\kfin\cdot\vec{\dum}} 
& = & \vec{\nabla}\Phi_f(\vec{\dum}) + \cdots. \label{eq:d4-81}
\end{eqnarray}
The gradient terms above correspond to the first term in the brackets of
Eq.~(\ref{eq:d4-2}); see also 
Eqs.~(\ref{eq:mainterm10}) and (\ref{eq:mainterm11}). 
In the low-energy limit, $\Phi_i(\vec{\dum}) \rightarrow \vec{\dum}\cdot\eps$,
so  $e\Phi$ is  responsible for the electric dipole interaction.

Since $\vec{A}$ contains both a $\Phi$ and a non-$\Phi$ term,
each $H^{\mathrm{int}}$ is the sum
of two terms and then there are four different possibilities for each
term of Eq.~(\ref{eq:d4-1}).  The largest term, which includes an
electric dipole $E1$ interaction at
both $N\gamma$ vertices, is the one for which both matrix elements contain
$\Phi$.  It is calculated in detail in Appendix A,
and certain aspects of it will also
be discussed in this section.  All terms in which
one of the matrix elements contain $\Phi$ are calculated
and tabulated in Appendix G of Karakowski's thesis\cite{thesis}.
These include all magnetic and spin-induced interactions occurring
at exactly one vertex.
Most of the terms that do not contain $\Phi$ are very small;
see  Appendix G of Ref.~\cite{thesis}.  

The largest second-order terms are the ones in which 
$-\int \vec{J}(\vec{\dum})\cdot \vec{\nabla}\Phi(\vec{\dum}) d^3\dum$
is substituted for all four $H^{\mathrm{int}}$'s in Eq.~(\ref{eq:d4-1}).
To avoid using  an explicit expression for $\vec{J}$,  
we integrate by parts and use current conservation:
\begin{eqnarray}
\vec{\nabla}\cdot\vec{J}(\vec{\dum})
 =  -\frac{i}{\hbar}\left[ H,\rho(\vec{\dum}) \right], \quad
\rho(\vec{\dum})  =  \sum_j e_j \delta(\vec{\dum} - {\vec{x}}_j).
\end{eqnarray} 
The integral over $\dum$ can then be performed
to give
\begin{eqnarray}
H^{\mathrm{int}} 
& = & -i [H, (e/\hbar) \Phi(\vec{x}_p) ],
\end{eqnarray}
where $H$ is now the full Hamiltonian.  If we examine only the piece of this commutator containing the 
proton kinetic energy, we see that
\begin{equation}
H^{\mathrm{int},p} =
 -i \left[\frac{p_p^2}{2m_p},  (e/\hbar) \Phi(\vec{x}_p) \right] =
-\frac{e}{m_p} \vec{p}_p \cdot \vec{A}(\vec{x}_p),
\end{equation}
so   the $\Phi$ commutators contain 
the operators responsible for E1 transitions.

We switch to the center-of-mass variables $\vec{p} \equiv (\vec{p}_p - \vec{p}_n)/2$ and
$\vec{P} \equiv \vec{p}_p +\vec{p}_n$ when we evaluate the commutator with
the full Hamiltonian.  Defining the ``internal'' Hamiltonian $H^{np} \equiv \frac{p^2}{2m_p} + V,$ 
we get for the $\Phi_i$ term 
\begin{equation}
H^{\mathrm{int}} 
 =  -\frac{e}{m_p}\vec{P}\cdot\eps e^{i\vec{k}_i\cdot\vec{r}/2} e^{i \vec{k}_i\cdot\vec{R}}  -i 
 \left[ H^{np} , (e/\hbar)\Phi_i(\vec{r}/2) \right]  
e^{i \vec{k}_i\cdot\vec{R}}.\label{eq:d4-3}
\end{equation}

The term with the $\vec{P}$ operator
will be called the ``recoil correction''; its matrix 
elements are calculated in Appendix B.   
The substitution 
$\vec{x}_p = \vec{r}/2 + \vec{R}$ generates
an exponential of the form $e^{i\vec{P}\cdot\vec{R}}$,
which  combines with
other similar exponentials to create the momentum-conserving delta function of
 Eq.~(\ref{eq:d3-6}).
 The net result of ignoring the recoil operators, therefore,  is to replace
$\vec{x}_p$ by $\vec{r}/2$ and switch from $\matrixt{}$ to $\matrixm{}$.

The interesting physics is contained in the second term of 
Eq.~(\ref{eq:d4-3}).
Plugging this into Eq.~(\ref{eq:d4-1}), we get
\begin{eqnarray}
\matrixm{a, \mathrm{uncr}}  & = &  -\sum_{C} \left\{\frac{ \mx{d_f}{[H^{np},
\hat{\Phi}_f] }{C}
\mx{C}{[H^{np}, \hat{\Phi}_i] }{d_i} }{\hbar\winit+E_{d_i}-E_C-
P_C^2/2m_d+i\varepsilon} \right.\nonumber - \\
& &  \left.\frac{ \mx{d_f}{[H^{np}, \hat{\Phi}_f] }{C}
\mx{C}{[H^{np}, \hat{\Phi}_i] }{d_i} }{\hbar\winit+E_{d_i}-E_C- P_C^2/2m_d+i\varepsilon} \right\}.
\end{eqnarray}
The dimensionless operators  $\hat{\Phi}_i \equiv (e/\hbar)\Phi_i(\vec{r}/2)$
and $\hat{\Phi}_f \equiv (e/\hbar)\Phi_f(\vec{r}/2)$  have been introduced,
and  this term has been
 labeled with the superscript ``$a$''.

Since $\ket{d_{i,f}}$ and $\ket{C}$ are eigenstates of $H^{np}$,
the commutators can be expanded and some cancellations with the
denominator can be made.  This generates four terms,
after adding the crossed term of Eq.~(\ref{eq:d4-1}):
\begin{equation}
\matrixm{a}  =  \matrixm{a1} + \matrixm{a2} + \matrixm{a3} + \matrixm{a4},
\end{equation}
\noindent where
\begin{eqnarray}
\matrixm{a1} & = & \left[ \frac{(\hbar\winit)^2}{2m_d} - \hbar\winit \right]^2 \sum_C
\frac{ \mx{d_f}{\hat{\Phi}_f}{C}
\mx{C}{ \hat{\Phi}_i}{d_i} }{\hbar\winit- \frac{(\hbar\winit)^2}{2m_d}+E_{d_i}-E_C+i\varepsilon}
, \label{eq:d4-4}  \\
\matrixm{a2} & = & \left[ \hbar\wfin + \frac{(\hbar\wfin)^2}{2m_d} \right]^2  \sum_C \label{eq:d4-5}
\frac{ \mx{d_f}{\hat{\Phi}_i}{C} \mx{C}{ \hat{\Phi}_f}{d_i}}{-\hbar\wfin- \frac{(\hbar\wfin)^2}{2m_d}+
E_{d_i}-E_C+i\varepsilon},  \\
\matrixm{a3} & = & \left[\hbar\wfin + \frac{(\hbar\wfin)^2}{2m_d} - 
\hbar\winit +  \frac{(\hbar\winit)^2}{2m_d} \right]
\mx{d_f}{\hat{\Phi}_f\hat{\Phi}_i}{d_i} \label{eq:d4-6},  \\
\matrixm{a4} & = & \frac{1}{2} \mx{d_f}{\left[ \ [H^{np},\hat{\Phi}_i],  \hat{\Phi}_f \right] +
\left[\ [ H^{np},\hat{\Phi}_f  ] ,  \hat{\Phi}_i \right] }{d_i} \label{eq:d4-7}.
\end{eqnarray}

The first two terms correspond to the Feynman diagram Figure~\ref{fig:d1-1}(b) and (c), and includes the case where both vertices
are E1 interactions.  The third term is a small recoil
correction, calculated in Appendix B.  
The final term is responsible for
compliance to the low-energy theorems  which result from
demanding gauge invariance; see   Section~\ref{sec:let}.

The evaluation of Eqs.~(\ref{eq:d4-4},\ref{eq:d4-5})
 while more complex than the one-body terms,
is straightforward except
for the additional complications of the intermediate
states and energy denominators.  
We can obtain a compact expression by first examining
the algebraic representation of the $\Phi$
operator in the appendix, equations~(\ref{eq:mainterm10})
and (\ref{eq:mainterm11}),
and noting  that it
contains distinct angular and radial pieces, which we call 
$O$ and $J$, respectively (suppressing the sums and related indices
here for simplicity).
Then we find  the $\matrixm{a1}$ term can be rewritten as 
\begin{eqnarray}
\matrixm{a1} & = & \sum_{\hat{C}} \sum_{l=0,2} \sum_{l'=0,2}
\mx{l'11M_f}{O_f}{\hat{C}} \mx{\hat{C}}{O_i}{l11M_i} \times\nonumber\\
& & \int_0^{\infty} \int_0^{\infty}  r  \chi_f^{l'\hat{C}}(r)
 J_i(r) u_l(r) dr, \label{eq:d4-8}
\end{eqnarray}
where the angular quantum numbers $(L_C S_C J_C M_C)$,
are denoted collectively by $\hat{C}$. The energy denominators have been
incorporated in  the function $\chi_f^{l'\hat{C}}(r)$.
All of the second-order terms can be reduced to the form
of Eq.~(\ref{eq:d4-8}).
We determine    $\chi_f^{l'\hat{C}}(r)$ by solving 
\begin{equation}
\left[ \frac{d^2}{dr^2} + \frac{m_p}{\hbar^2}\left[ E_0 -
V_{\hat{C}}(r) \right]- \frac{L_C(L_C+1)}{r^2}\right] r \chi_f^{l'\hat{C}}(r)
=
\frac{m_p}{\hbar^2}\frac{u_{l'} (r) J_f(r)}{r} \label{eq:d4-10}.
\end{equation}
where
 $E_0 \equiv \hbar\winit
 - \frac{(\hbar\winit)^2}{2m_d}+E_{d_i}.$
 For $V_{\hat{C}}(r) $, we use the Reid93 $np$ potential\cite{sto94}.
This is an ordinary, inhomogeneous, second-order differential equation.
 The boundary condition is that $ \chi_f^{l'\hat{C}}(r)$ 
must be an outgoing spherical wave at large distances.

\subsection{Relativistic Corrections \label{sec:rel} }

The major relativistic correction here is the spin-orbit effect. Corrections
arising from boosting the final deuteron wavefunction into  a moving frame
are  small here and neglected.
The origin of the relativistic spin-orbit effect
can be understood by examining  
the magnetic term in the Hamiltonian of Eq.~(\ref{eq:d2-3}):
\begin{equation}
H^M =
\sum_{j=n,p} \frac{e(1+\kappa_j)}{2m_j} \vec{\sigma}_j\cdot\vec{B}(\vec{x}_j). 
\end{equation}
The relativistic correction to the magnetic
field for an object moving with velocity $\vec{v}$ is
$
\vec{B} \rightarrow \vec{B} - \vec{v}\times\vec{E},
$
which leads to  a relativistic term in the Hamiltonian:
\begin{equation}
H^{RC} = -\sum_{j=n,p}
\frac{e_j(1+2\kappa_j)}{4m_j}
\vec{\sigma}_j\cdot\left(\vec{v}_j\times\vec{E}_j\right).
\end{equation}
Only the proton gives a correction
because of the factor $e_j$.
The effects of Thomas precession are  included.
Since $\vec{v} \approx \vec{p}/m$ and  
$\vec{E} = \vec{\nabla}V(r) \sim \vec{r}$
for a central potential, the dot product above goes as 
$\vec{\sigma}\cdot\vec{L}$, where $\vec{L}$ is the orbital angular momentum. 
If we make the minimal  substitution
$ \vec{p} \rightarrow \vec{p} - e\vec{A}, $                              
and use
$
\vec{E} = -\frac{\partial\vec{A}}{\partial t} ,
H^{RC}$ becomes
\begin{equation}
H^{RC} =\sum_{j=n,p} \frac{e(1+2\kappa_j)}{4m_j^2} e_j \vec{\sigma}_j\cdot
\left[ \vec{A}(\vec{x}_j) \times
\frac{\partial}{\partial t} \vec{A}(\vec{x}_j)  \right], 
\end{equation}
and the resulting  correction to the scattering amplitude is given by 
\begin{equation}
\matrixm{RC} = \bra{d_f} \frac{-ie^2}{4m_p^2}
\vec{\sigma}_p\cdot(\epspr\times\eps)
 (\hbar\wfin + \hbar\winit ) (2\kappa_p+1)
 e^{-i\vec{q}\cdot\vec{r}/2} \ket{d_i}. \label{eq:d5-1}
\end{equation}

\subsection{Checking the Calculation \label{sec:let} }

There are checks which can be performed to help ensure that certain
parts of the calculation are correct.  One  involves the use of
the optical theorem to calculate the total cross-section
$\sigma_{\mathrm{tot}} $for deuteron 
photodisintegration, $\gamma d\to n p$.
With our normalization conventions, the
optical theorem is given by
\begin{equation}
\sigma_{\mathrm{tot}}  =  \frac{4\pi}{\omega} \imag \matrixm{}(\theta = 0),
\end{equation}
in which $ \matrixm{}$ is our Compton amplitude.
 Only the diagram  of  Fig.~1b  enters      here.   
However, all combinations of interactions at the two
vertices must be included.  In a 1964 paper, 
Partovi \cite{pa64} calculated the contributions from
each of these interactions at 3 different energies.  
The dominant term is the one in which
there is an E1 interaction at each vertex.  He called
this cross-section ``approximation A'', and added the 
other terms one by one in successive approximations, until
every term was included (``approximation I'').
These results are reproduced in Table~\ref{t:d6-1}, along
with the corresponding
cross-sections extracted from this calculation.  There is
reasonable agreement at all energies;
any differences can be attributed to
changes in the short-range parts of the  potentials. 
The meanings of the approximations in Table~\ref{t:d6-1}:
Approximation A: E1 only;
Approximation B: A + singlet M1;
Approximation C: B + E2; Approximation D: C + triplet M1;
Approximation E: D + spin induced triplet M1;
Approximation F: E + spin induced triplet M2;
Approximation G: F + spin induced triplet E1;
Approximation H: G + retardation corrections to E1;
Approximation I: H + all other terms.


We also compare these
photodisintegration cross-sections with experimental data up to 100 MeV.
This is shown in Fig.~\ref{f:d6-5}.  
A vertex correction (see Fig.~1f) that
contributes to this process, and is responsible for
dominant meson exchange correction\cite{br} is  also  included in a
second graph of the range 50--100 MeV.
There is good agreement with the data.  Our
photodisintegration cross-section also reproduces the
correct threshold behavior, including pion-exchange current effect,
as described by Brown \&Jackson\cite{br}.

Another important check is to ensure
that the calculation is gauge invariant.
Since the relativistic two-photon scattering amplitudes 
must have
the form $\epsilon_i^{\mu}T_{\mu\nu}\epsilon_f^{\nu}$,
 the condition imposed 
by gauge invariance is that 
$
k_i^{\mu}T_{\mu\nu}\epsilon_f^{\nu} = \epsilon_i^{\mu}T_{\mu\nu} k_f^{\nu} = 0.
$
An important
consequence of this is what is usually referred to as the low-energy theorem.
Friar \cite{fr75} showed,
assuming only gauge invariance,  that
 the full
Compton amplitude must go as 
\begin{equation}
{\mathcal M} \rightarrow
\frac{e^2}{m_d} (\hat{\epsilon}_f\cdot\hat{\epsilon}_i).
\end{equation}
as the photon energy goes to zero.
This result  is 
non-trivial because it involves the target(deuteron) mass; at low-energy
the seagull amplitude depends on the \emph{proton} mass :
\begin{equation}
{\mathcal M}_{\mathrm{sg}} \rightarrow \frac{e^2}{m_p}
(\hat{\epsilon}_f\cdot\hat{\epsilon}_i).\label{newamp}
\end{equation}
Therefore, all other terms, including meson-exchange terms,
must exactly cancel half of the seagull
amplitude at threshold. That 
this occurs is easy to show, 
if we ignore meson-exchange terms.
The only term which survives in the low-energy
limit, from Eq.~(\ref{eq:d4-7}), is
\begin{equation}
{\mathcal M}^{a4}
=\frac{1}{2} \mx{d_f}
{\left[ [ \frac{p^2}{m_p} ,\hat{\Phi}_i],  \hat{\Phi}_f \right] +
\left[\ [ \frac{p^2}{m_p},\hat{\Phi}_f  ] ,  \hat{\Phi}_i \right] }{d_i},
\label{eq:d6-3}
\end{equation}
where we have only included the kinetic energy part of $H^{np}$;
the potential energy term is cancelled by 
meson-exchange currents and will be discussed shortly.
The amplitude of Eq.~(\ref{newamp})
 can easily be shown to satisfy Friar's low-energy theorem.
Since $\vec{A} \rightarrow \hat{\epsilon}$ at threshold (the 
dipole approximation), the definition of $\hat{\Phi}_i$
Eqs.~(\ref{eq:d4-80}) and (\ref{eq:d4-81}) implies that 
$
\hat{\Phi}_i  \rightarrow
\frac{e}{\hbar} \frac{\vec{r}}{2}\cdot\hat{\epsilon}_i, \quad
\hat{\Phi}_f  \rightarrow
\frac{e}{\hbar} \frac{\vec{r}}{2}\cdot\hat{\epsilon}_f. 
$
Evaluating the commutators of Eq.~(\ref{eq:d6-3}) then yields
\begin{equation}
{\mathcal M}^{a4}
\rightarrow -\frac{e^2}{2m_p} (\hat{\epsilon}_f\cdot\hat{\epsilon}_i),
\end{equation}
which cancels half of the seagull term as required.  
Our numerical calculation
reproduces this result.
The differential cross-section in
this figure includes all terms except meson-exchange terms,
and is seen to approach the correct limiting value.  
This plot shows only the forward cross-section,
but the low energy theorem is satisfied at all angles. 

We now examine the effects of the meson-exchange
currents that were previously neglected.
Since the potential energy portion of the double
commutator term (Eq.~\ref{eq:d6-3})
has not yet been accounted for, we suspect that
this must cancel the explicit meson-exchange
terms at low energy.  This is indeed the case, as shown by
Arenh{\"{o}}vel \cite{ar80}.
Fig.~\ref{fig:d6-2}
shows the four pion-exchange diagrams 
which contribute         and 
cancel the double-commutator $V^{\mathrm{OPEP}}$ term at threshold. 
These four terms and the
double-commutator $V^{\mathrm{OPEP}}$ term are described in detail in
 in App.~I of \cite{thesis}.  
Fig.~\ref{fig:d6-15} shows a graph of the  differential
cross-section at $0^{\circ}$ which includes only these 5 terms.
It is exactly zero at threshold.  This is true at all angles, not
just in the forward direction.
Gauge invariance has been satisfied without explicitly demanding
it in the formulation of the problem.   
The cross-section can be seen to have an $\omega^4$ dependence,
which means that the amplitude goes like $\omega^2$.

\section{Results and Discussion \label{ch:res} }
We now compute  the deuteron
Compton scattering cross-sections at the three experimentally
relevant  three energies of 
49 MeV, 69 MeV  and 95 MeV.
\subsection{Effects of Various Terms in Cross-Section}

The first step is to
 study the contributions to the cross-section
from the major interactions.  They will be labeled as
follows: SG:  seagull term, 
EM: All interactions, except for recoil corrections, 
  arising from the multipole
decomposition of the vector potential as described in App.~A here 
 and  App.~G of \cite{thesis}.
These include all electric and magnetic dipoles,
as well as
the double commutator term needed to satisfy the low-energy theorem.
$\pi$:  All pion interactions pictured in Fig~\ref{fig:d6-2}
which are needed to satisfy the low-energy theorem.
These are detailed  in App.~J of \cite{thesis}.
$\alpha + \beta$:  polarizability terms.
Unless
otherwise specified,
the values
for the polarizabilities
are chosen as 
$\alpha_p = 10.9$, $\alpha_n = 12.0$, $\beta_p = 3.3$, $\beta_n = 2.0$,
as determined in \cite{fe91,sh91}.
    These major terms  MT are  displayed  in 
 Fig.~5 and on the top of Fig.~6.  
The data from  \cite{lu94} is displayed on
all graphs at 49  and 69 MeV; no  data have yet  been published for 95
MeV. The seagull  term alone  provides a  reasonable description of the
data at 49 MeV,  and does no  worse than all the  terms together at 69
MeV. At  low  energy, the  seagull term  has a  $(1  +\cos^2{\theta})$
dependence.   As  the  energy   becomes   higher,  this  dependence is
maintained   at the   forward  angles,  while  the   cross-section get
increasingly smaller  at the backward angles.  It is the dominant term
at all energies under  consideration. The EM  terms raise the  seagull
cross-section and become   increasingly important as the photon energy
rises.  The $\pi$ terms,  on the  other hand, do not have a very large
effect, even at  95 MeV. These  terms exactly  cancel at threshold and
continue to nearly cancel at  higher energies. 

These terms are collectively called the Major Terms (MT).  They describe the
data reasonably well.
The smaller  corrections are:
RC: The Relativistic Correction.
CMC: Recoil Corrections (from the  Center-of-Mass operator),
as calculated in Appendix B.  This 
also includes only the largest correction.
VC:  $\gamma N$ Vertex Corrections, such as the one pictured
in Fig.~\ref{fig:d1-1}(f).  These are detailed in \cite{thesis}.
These terms produce significant effects only at 95 MeV.
A detailed
 examination of  the EM shows that  multipoles
higher than $L=1$ are unimportant below 100 MeV. 



\subsection{Determining Polarizabilities from Deuteron Cross-Sections}
Having displayed       the major contributions, we now investigate
the dependence of the
cross-section on the size of the neutron polarizabilities.  
Figs.~\ref{fig:e2-1}-~\ref{fig:e2-2}
show the effects of varying  the magnitudes of $\alpha_n$ and $\beta_n$.
The $\alpha_n$ term has the same $(1+ \cos^2{\theta})$ dependence
as the seagull term at low energies,
while the $\beta_n$ term goes like $\cos{\theta}$, so that 
 the magnetic polarizability term gives no contribution at $90^{\circ}$.
The largest effects for both terms occur
at extreme forward and backward angles, while
the experimental data points are in the middle.
Figs.~\ref{fig:e2-1} and 8 show the changes to the cross-section over the 
range of $\alpha_n = 12.0 \pm 4.0$ and $\beta_n = 2.0 \pm 4.0$, which is
a range of variation that is
slightly larger than the errors quoted in \cite{sh91}.
These  illustrate the sensitivity to
the size error.  At  49 MeV the effects of varying the
polarizabilities
between the values of $\alpha_n = 8.0 - 12.0$ and  $\beta_n = 2.0 - 6.0$, are
very small 
compared with  the experimental
error bars. 
There are only two data points at 69 MeV,
and one of these lies above our  calculated cross-section.  
The data point at the greatest angle is also difficult to describe
at 49 MeV.  Other calculations show
similar inconsistencies with these two points \cite{ar95,lv98,ch98}.
Looking at (Fig.~\ref{fig:e2-2}), 
we see that using a lower $\alpha_n$ 
and a higher $\beta_n$ than the accepted values provide the best
description of
 these points.

 We must keep in mind that the proton polarizabilities also have
uncertainties, and they contribute to the deuteron cross-section just
as much as the neutron.  Another set of graphs as described above is
given for $\alpha_p$ and $\beta_p$ (Figs.~\ref{fig:e2-3} and 11)
,  using only the ranges as the actual
experimental error given in \cite{fe91}.   The same trends stand out
here, most notably that a smaller value of $\alpha$ and a larger value
of $\beta$  would provide a better description of the data points at
the back angles. Using the uncertainties
in neutron and proton polarizabilities
would allow a perfect reproduction of the data.

 The chance to  determine  polarizabilities
 looks more promising at 95 MeV, as shown in Figs.~8 and 11.
 The effects of the
polarizabilities  are larger, and the range of $\pm 4.0$ probably will
extend beyond the experimental error bars at angles such as
$50^{\circ}$ and $140^{\circ}$.    







\subsection{Sources of Error}
We would like to study the effects of some of the smaller terms more closely.
For each term we estimate the absolute error
in the value $\alpha_n$ (we fix $\beta_n = 2.0$ in this section) caused
by neglecting various effects.  These are listed in Table~\ref{t:e3-1}
and shown in some  figures.
First, we examine the gauge invariant set of pion terms.  
The results in  Table~\ref{t:e3-1} show   that the value $\alpha_n$
would be slightly underestimated  at the forward angles
and overestimated at backward angles, and would not be affected
around $120^{\circ}$.   The magnitude of this error
is about the same as the experimental error in the polarizability,
so these effects  need to be included 
to make an accurate determination of  $\alpha_n$.
The effect is same at all energies because 
both the $\pi$ terms and polarizability terms have an $\omega^2$ dependence.
The recoil (CMC) corrections would be difficult to see on any graph,
and this is also shown in the table.  The largest errors
are actually at the lowest energies, indicating that
these terms are of order $\omega^1$.
The   small values for $\Delta\alpha_n$
at 95 MeV 
indicate that these terms
can be neglected here.

The relativistic corrections (RC) (Fig.~\ref{fig:e3-12})
are not a "small" correction: 
even at 49 MeV they would have to be included in a calculation
to make an accurate determination of  $\alpha_n$.  The amount of
error that would be introduced by neglecting
these terms is comparable to $\alpha_n$ itself at some angles.

We turn to the vertex corrections (VC) of Fig.~1f which
are  also necessary  for an accurate estimate of the polarizability.
Their effect is
greater than the effects of the other pion terms combined.
The errors decrease as energy increases, but even
at 95 MeV the errors, particularly at middle angles, is at least as
large as previous errors in polarizability determinations.  
At backward angles the effect seems to be independent of energy. See 
 Fig.~\ref{fig:e3-13}.
Finally, we note that the effect
 of the uncertainly in the $\pi N$ coupling constant $f$ on computed values of
 the deuteron Compton scattering are very small.


\subsection{Tensor-Polarized Deuteron Compton Scattering}

Another interesting observable is the
tensor-polarized deuteron Compton scattering cross-section. Chen\cite{chen98}
introduced the idea that the quantity
\begin{equation}
\left( \frac{d\sigma}{d\Omega} \right)_T
= \frac{1}{4} \left[ 2 \frac{d\sigma}{d\Omega} (M_i = 0) - 
\frac{d\sigma}{d\Omega} (M_i = 1) - \frac{d\sigma}{d\Omega} (M_i = -1) \right]
\end{equation}
be studied because it isolates the importance of pion exchange terms in
effective field theory treatments of deuteron Compton scattering.
Fig.~\ref{fig:e4-1} shows that $ \left( \frac{d\sigma}{d\Omega} \right)_T$,
as obtained in our complete calculation 
has only
a slight dependence on the polarizabilities, which are our primary interest.
We
display our results to facilitate future
comparisons between potential-model and
effective field theory calculations. In our approach, the
contributions of magnetic interactions (MI) and
relativistic corrections (RC) are the dominant contributions to
$ \left( \frac{d\sigma}{d\Omega} \right)_T$. 
We  define
the pion-exchange terms to  include any pions that
give rise to the tensor force (and 
the $l=2$ state)  in  the deuteron.
However,
to be more consistent with the calculation of
\cite{chen98} we now will include only one pion exchange in the pion terms
and no pion exchanges
in the magnetic terms.
Therefore, the $D$-state deuteron
wavefunction $u_2(r)$ will be set to zero in all
terms except for the Thomson term
and the $D$-state contributions to this term will be grouped
with the pion-exchange terms.
Defined in this way, the pion-exchange effects
turn out to be 
 less than about 0.5 nb/sr at all angles.   
The results for the
magnetic interactions (MI)  are shown in Fig.~\ref{fig:e4-2}.
These interactions include all of the dispersive terms of
Eq.~(\ref{eq:mainterm1}) as well as the
seagull term of Eq.~(\ref{sgt})
  computed  without the deuteron $D$-state.    
In addition, we show the effect of including the
deuteron $D$-state in the calculation of the magnetic terms (but not the
seagull term).  This is a non-negligible effect.
Finally, 
the relativistic correction (RC) is added (including the $D$-state).
Since this is a spin-orbit effect,
we group this with the magnetic interactions.
Both the relativistic term and the deuteron $D$-states have a large
effect in the calculation;  this observable is very
sensitive to these individual effects.


\section{Conclusion \label{ch:conc} }
We have presented a calculation of deuteron
Compton scattering valid for energies less than 100 MeV.
All terms that we believe to be important are included.
  The seagull, lowest-order
multipoles, and
polarizability effects dominate the cross-section.
Relativistic corrections must also 
be incorporated in the calculation.
Although smaller than the preceding terms, pion-exchange and 
vertex corrections are also needed to accurately determine the
polarizabilities.  We conclude that
recoil effects and multipoles of second-order and higher are
negligible in this energy range. We are able to achieve a reasonable
agreement with the existing data; see Figs.~5,7 and 10.

This is the most extensive study so far of the effects of
various terms in the cross-section on the value of $\alpha_n$; see Tables III
and IV.
However, certain corrections which are believed to be small
have been omitted.  No contributions from 
the $\Delta$ resonance have been explicitly included,
although the magnetic polarizability itself is thought
to contain a
large paramagnetic contribution from the $\Delta$.
The $\Delta$ resonance has been shown to
have a small
effect
on deuteron
photodisintegration at energies less than 100 MeV \cite{yi88}.
In addition, considering the rather small contributions
from $\pi$-exchange currents,
heavier mesons such as the $\rho$ have also been neglected.  
Higher-order relativistic
corrections have not been included, but may need to be investigated further
at energies
near 100 MeV because of the large effects of the leading-order terms there.

The only fully detailed
calculation of this type that has been published to date was by
Weyrauch \cite{we88,we90}.
There are several important differences between his
calculation and the present one.  
One of these is the $NN$ potential used.  He used  a
separable $NN$ T matrix, while here
we have found the Green's function for an intermediate $NN$
state interacting via the Reid93 potential.
He also assumes meson-exchange currents  and relativistic corrections
to be negligible.  Probably the most important
difference, however,  is in gauge invariance.  Gauge invariance is
broken in the separable potential model and 
needs to be artificially restored.  We have made no such assumptions
here.  Gauge invariance is a natural consequence
of including all of the appropriate terms.

The cross-section predicted by Weyrauch is higher than the experimental
data.  However, recent 
results of Arenh{\"{o}}vel \cite{ar95}
and Levchuk and L'vov
\cite{lv95,lv98} are in qualitative agreement with the data.  Their work is
similar to ours in that they also take 
 diagrammatic approach using  realistic
potentials and including 
meson-exchange currents. The full details of their calculations
are  not yet  available, so precise comparisons are not easily made.
Lastly, a new calculation using effective field theories is  also
in qualitative agreement with  the data \cite{ch98}.
We have also  followed \cite{chen98} in studying
tensor-polarized Compton scattering. In our formulation
the effects of the deuteron D-state and the relativistic correction
are the dominant contribution.

We return to our
 basic question of the feasibility of making an accurate determination 
of the neutron polarizability from a deuteron Compton scattering experiment. 
The available
data at 49 and 69 MeV
show that the ranges $\alpha_n = 12.0 \pm 4.0$ and $\beta_n = 2.0 \pm 4.0$
are reasonable,
but a more accurate estimate is difficult since the polarizability terms are
small relative to the experimental error bars.  The uncertainties in the
proton 
polarizabilities must also be taken into account.  These
problems are compounded by the data point at the greatest angle at each 
energy,
which we all authors find
difficult to
describe. 
The closest description of  all of the experimental data comes from lowering 
the value of $\alpha_n$ and
raising $\beta_n$.

Perhaps the most promising
estimates of the polarizability will come from data at higher energies.
Assuming that
the experimental error bars are not significantly larger,
we should find that the range of polarizabilities that fit the
data is smaller.  Sensitivity to both $\alpha$ and $\beta$ is
largest at the forward and backward angles, 
while the cross-section at $90^{\circ}$ is the
least sensitive to the polarizabilities.
Many of the ``small'' corrections, however,
are particularly small at $120-140^{\circ}$,
and the cross-section is still very sensitive
to the polarizabilities at these angles.   The
proton polarizabilities, unfortunately, will also make a larger contribution
to the cross-section, and there is no way to determine
the neutron polarizabilities more accurately than those of the 
proton in this experiment.   This energy is at the
limit of the validity of the low-energy expansion, which
may introduce errors as well.  In any case, studies
at higher energies should provide additional information 
on the polarizabilities.



\begin{table}
\caption{Experimental results for $\agen_p$ and $\bgen_p$ \label{t:c3-1} }
\begin{center}
\begin{tabular}{ cccc }
& \underline{$\omega_i$(MeV)} & \underline{$\theta$(degrees)} & \underline{result} \\
Moscow(1975) & 70-110 & 90,150 & $\agen_p + \bgen_p = 5.8 \pm 3.3$ \\
        & & & $ \agen_p - \bgen_p  = 17.8 \pm 2.0$ \\
Illinois (1991) & 32-72 & 60,135 &  $\agen_p = 10.9 \pm 2.2 \pm 1.3$ \\
& & & $ \bgen_p = 3.3 \mp 2.2 \mp 1.3$\\
Mainz(1992) & 98, 132 & 180 & $ \agen_p - \bgen_p = 7.03^{+2.49+2.14}_{-2.37-2.05} $\\
& & & $\agen_p  = 10.62^{ +1.25+1.07}_{-1.19-1.03}$\\
& & & $\bgen_p = 3.58^{ +1.25+1.07}_{-1.19-1.03}$\\
Illinois/SAL(1995) & 70-148 & 90,135 & $\agen_p = 12.5 \pm 0.6 \pm 0.7 \pm 0.5$  \\
& & & $\bgen_p = 1.7 \mp 0.6 \mp 0.7 \mp 0.5$ \\
\end{tabular}
\end{center}
\end{table}
\begin{table}
  \caption[Comparison of photodisintegration cross-sections calculated in several approximations]{ \label{t:d6-1} 
Comparison of photodisintegration cross-sections calculated
in several approximations.  $\sigma_P$ is the cross-section of
Partovi \cite{pa64}, while $\sigma_K$
is from this calculation.}

\begin{tabular}{ccccccccc}
        \multicolumn{1}{c}{Approximation} &
        \multicolumn{2}{c}{20 MeV} &
        \multicolumn{1}{c}{ } &
        \multicolumn{2}{c}{80 MeV} &
        \multicolumn{1}{c}{ } &
        \multicolumn{2}{c}{140 MeV} \\ 

&  $\sigma_{P}$  ($\mu b/sr$) & $\sigma_{K}$ ($\mu b/sr$) & 
&  $\sigma_{P}$  ($\mu b/sr$) & $\sigma_{K}$ ($\mu b/sr$) &
&  $\sigma_{P}$  ($\mu b/sr$) & $\sigma_{K}$ ($\mu b/sr$) \\ 
A & 579.1 & 583.3 & & 77.15 & 80.54 & & 34.04 & 34.56 \\
B & 589.2 & 593.2 & & 83.50 & 85.83 & & 39.77 & 38.59 \\ 
C & 591.5 & 595.0 & & 84.55 & 86.86 & & 40.38 & 38.99 \\  
D & 591.7 & 595.1 & & 84.85 & 87.06 & & 40.67 & 39.16 \\ 
E & 591.6 & 594.7 & & 84.73 & 86.61 & & 40.55 & 38.70 \\ 
F & 592.1 & 595.3 & & 85.43 & 87.68 & & 41.31 & 40.26 \\ 
G & 591.9 & 594.6 & & 90.45 & 90.22 & & 47.07 & 43.33 \\ 
H & 588.2 & 591.2 & & 87.52 & 86.36 & & 44.64 & 39.44 \\ 
I & 588.2 & 591.2 & & 87.40 & 86.40 & & 44.53 & 39.52 \\

\end{tabular}
\end{table}
\begin{table}
\caption{ \label{t:e3-1} Error in electric polarizability
  generated by omitting the gauge-invariant set of pion terms $\pi$ and
  the recoil corrections (CMC)} 
\begin{tabular}{c|ccc|ccc}
        \multicolumn{1}{r}{}&
        \multicolumn{3}{c}{$\pi$}&
        \multicolumn{3}{c}{CMC}\\
\underline{$\theta$(deg)} & \underline{49 MeV} & \underline{69 MeV} & \underline{95 MeV} &
\underline{49 MeV} & \underline{69 MeV} & \underline{95 MeV} \\ 
&&&&&&\\
0 & -2.9 & -2.6 & -2.0 & 0.0 & 0.0 & 0.0  \\
30 & -2.7 & -2.4 & -1.9  & -0.5 & -0.2 & -0.1  \\
60 & -2.1 & -2.0 & -1.6 & -1.1 & -0.5 & -0.2  \\
90 & -1.2 & -1.4 & -1.7 & 0.1 & 0.2 & 0.3   \\
120 & 0.0 & -0.3 & -0.9 & 1.4 & 0.8 & 0.5  \\
150 & 1.1 & 0.9 & 0.5 & 0.6 & 0.3 & 0.2 \\
180 & 1.5 & 1.4 & 1.1  & 0.0 & 0.0 & 0.0 \\

\end{tabular}

\end{table}

\begin{table}
\caption{Error in electric polarizability
  generated by omitting the relativistic correction (RC) and vertex correction
  (VC)} 
\begin{tabular}{c|ccc|ccc}
        \multicolumn{1}{r}{}&
        \multicolumn{3}{c}{RC}&
        \multicolumn{3}{c}{VC} \\
\underline{$\theta$(deg)} & \underline{49 MeV} & \underline{69 MeV} & \underline{95 MeV} &
\underline{49 MeV} & \underline{69 MeV} & \underline{95 MeV} \\
&&&&&&\\
0 & -7.9 & -9.2 & -11.9 & 3.7 & 1.4 & 0.9  \\
30 & -8.7 & -10.2 & -13.2 & 3.9 & 1.9 & 1.7  \\
60  & -9.7 & -11.8 & -15.8 & 3.7 & 2.7 & 4.1  \\
90   & -4.9 & -7.1 & -10.1 & -1.1 & 1.2 & 5.0   \\
120   & 5.1 & 4.1 & 3.4 & -8.0 & -3.4 & 0.9  \\
150  & 10.0 & 9.8 & 10.3 & -10.6 & -5.9 & -2.4  \\
180  & 11.1 & 11.1 & 12.0 & -11.1 & -6.6 & -3.3  \\
\end{tabular}

\end{table}
\begin{figure}

\caption{Feynman diagrams \label{fig:d1-1} }
\end{figure}
\begin{figure}
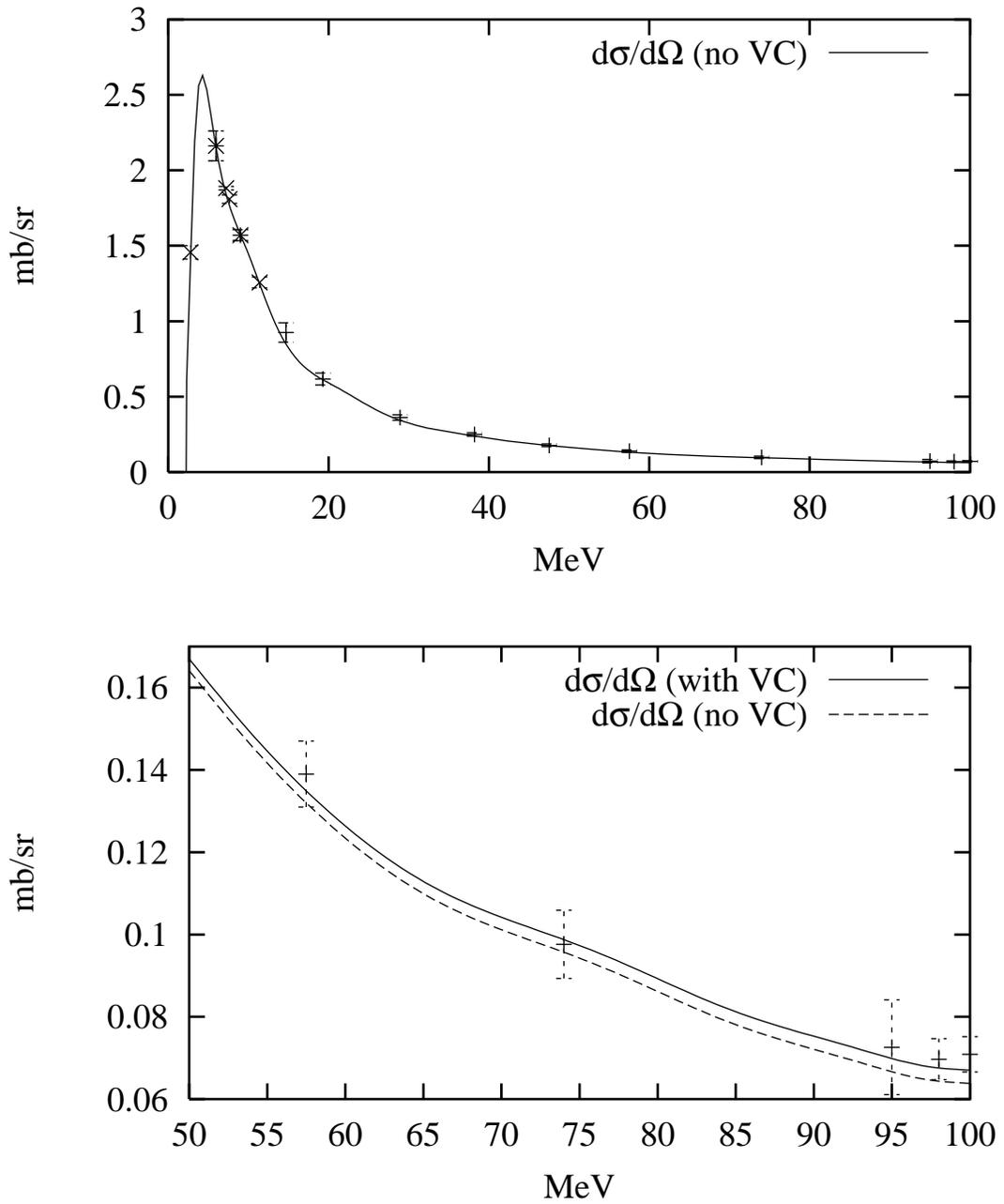

\centering
\caption{ \label{f:d6-5}
  Total photodisintegration cross-sections compared with experimental data.
}
\end{figure}
\begin{figure}
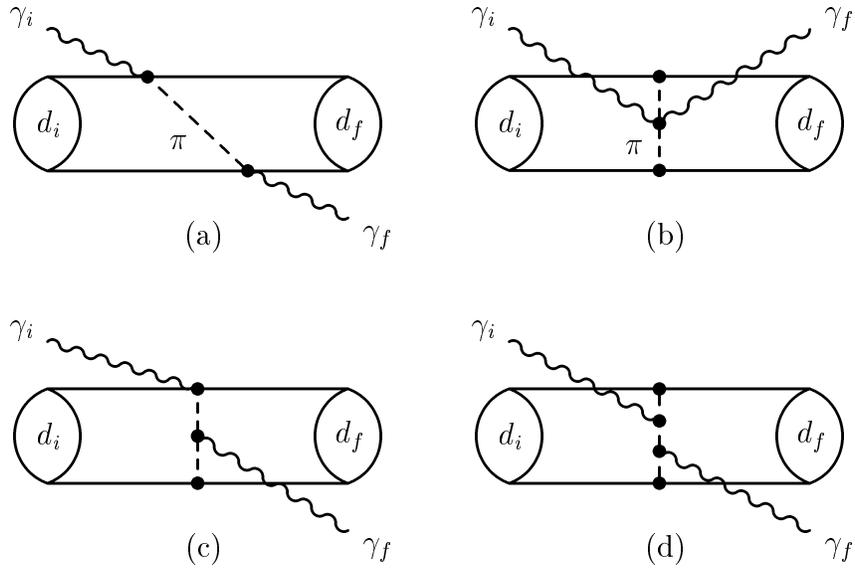

\caption{\label{fig:d6-2}
Gauge invariant set of pion-exchange diagrams}
\end{figure}
\begin{figure}
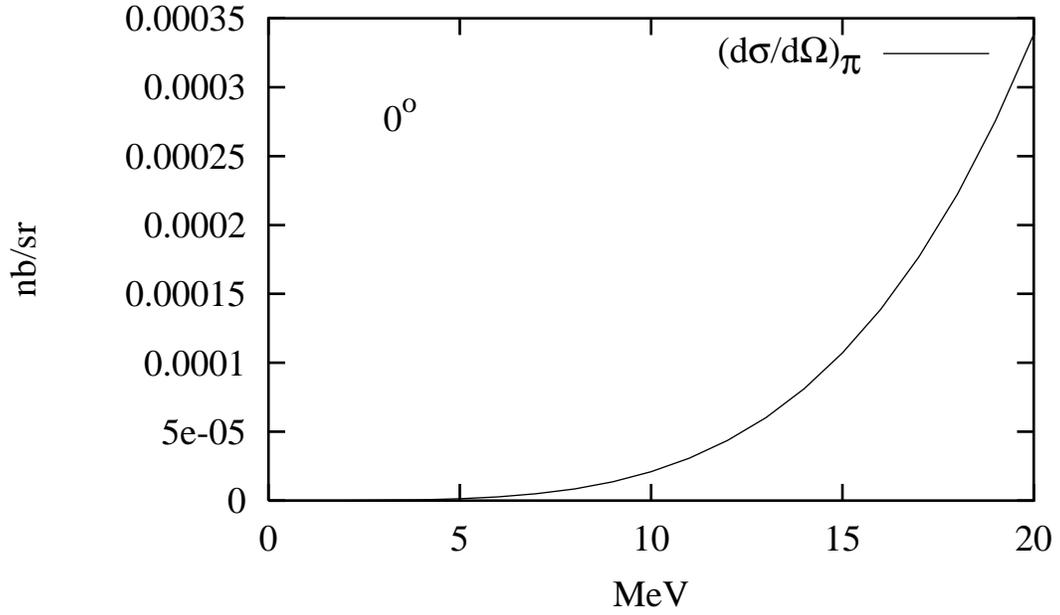

\centering
\caption{ \label{fig:d6-15} Forward
  differential cross-section
  including only pion-exchange and double-commutator
  $V^{\mathrm{OPEP}}$ contributions }
\end{figure}
\begin{figure}
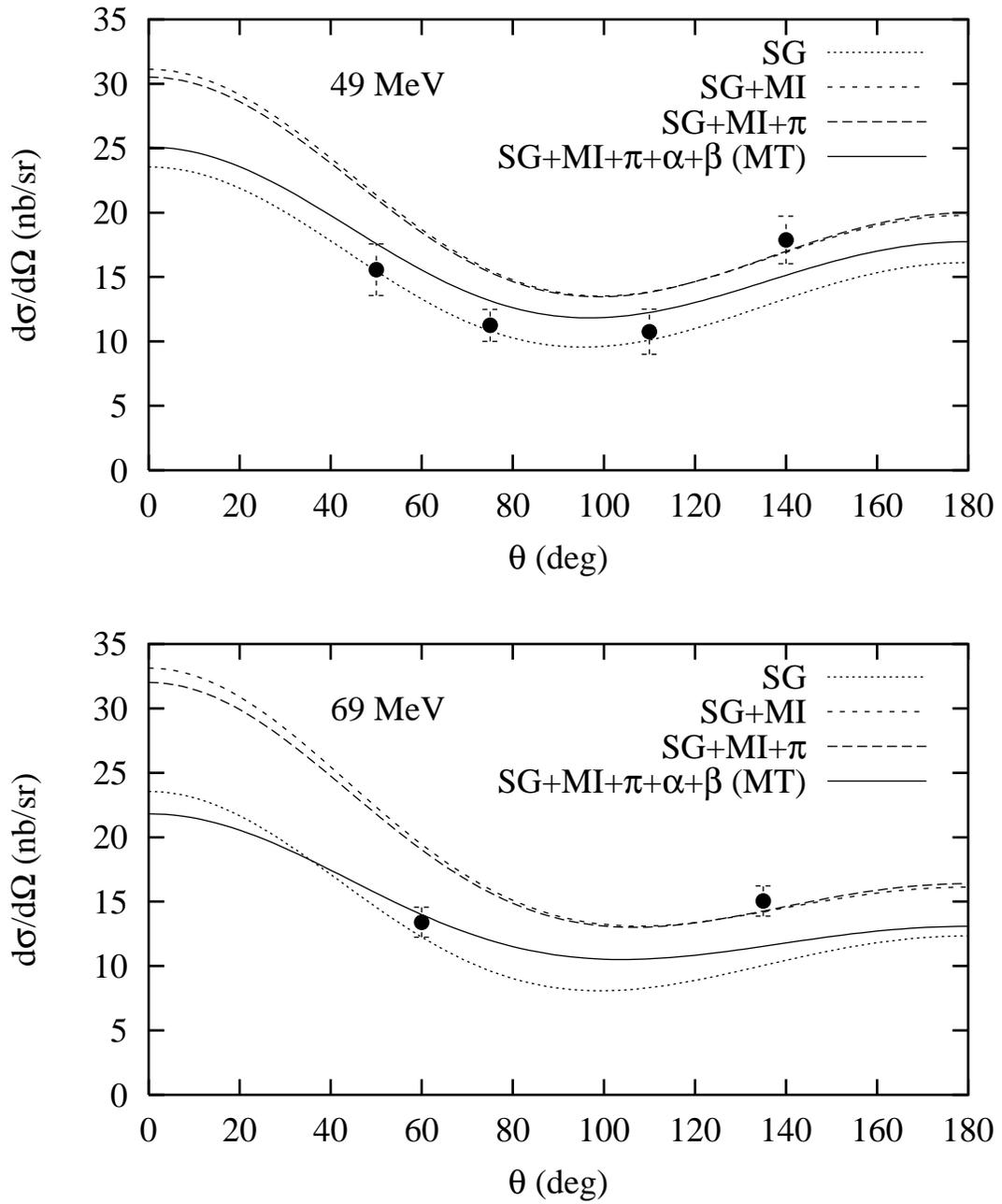

\centering
\caption{ \label{fig:e1-1} Contributions
  of different terms.} 

\end{figure}
\begin{figure}
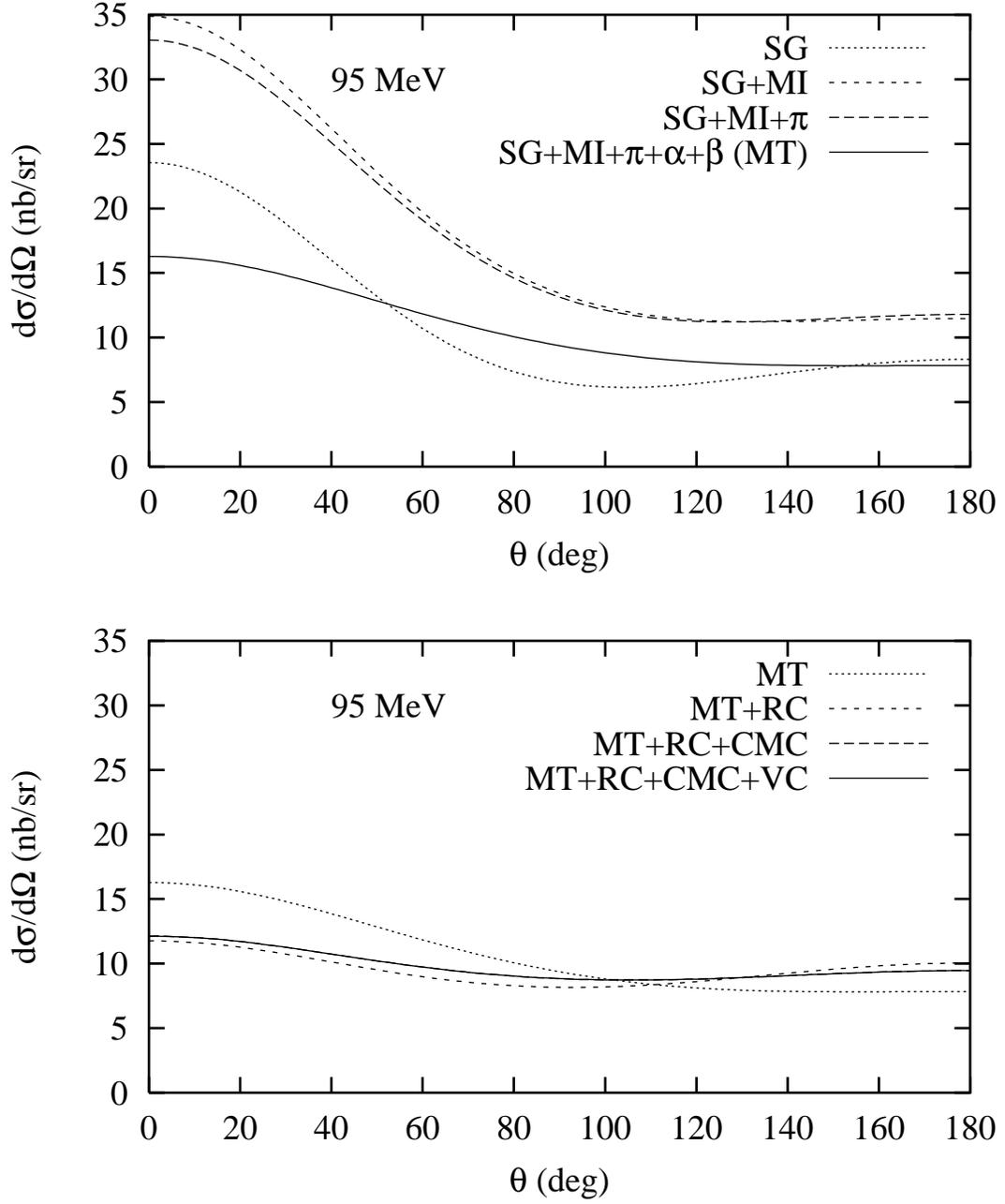

\centering
\caption{ Contributions
  of different terms at 95 MeV. Both major terms and smaller corrections
  are displayed} 
\end{figure}


\begin{figure}
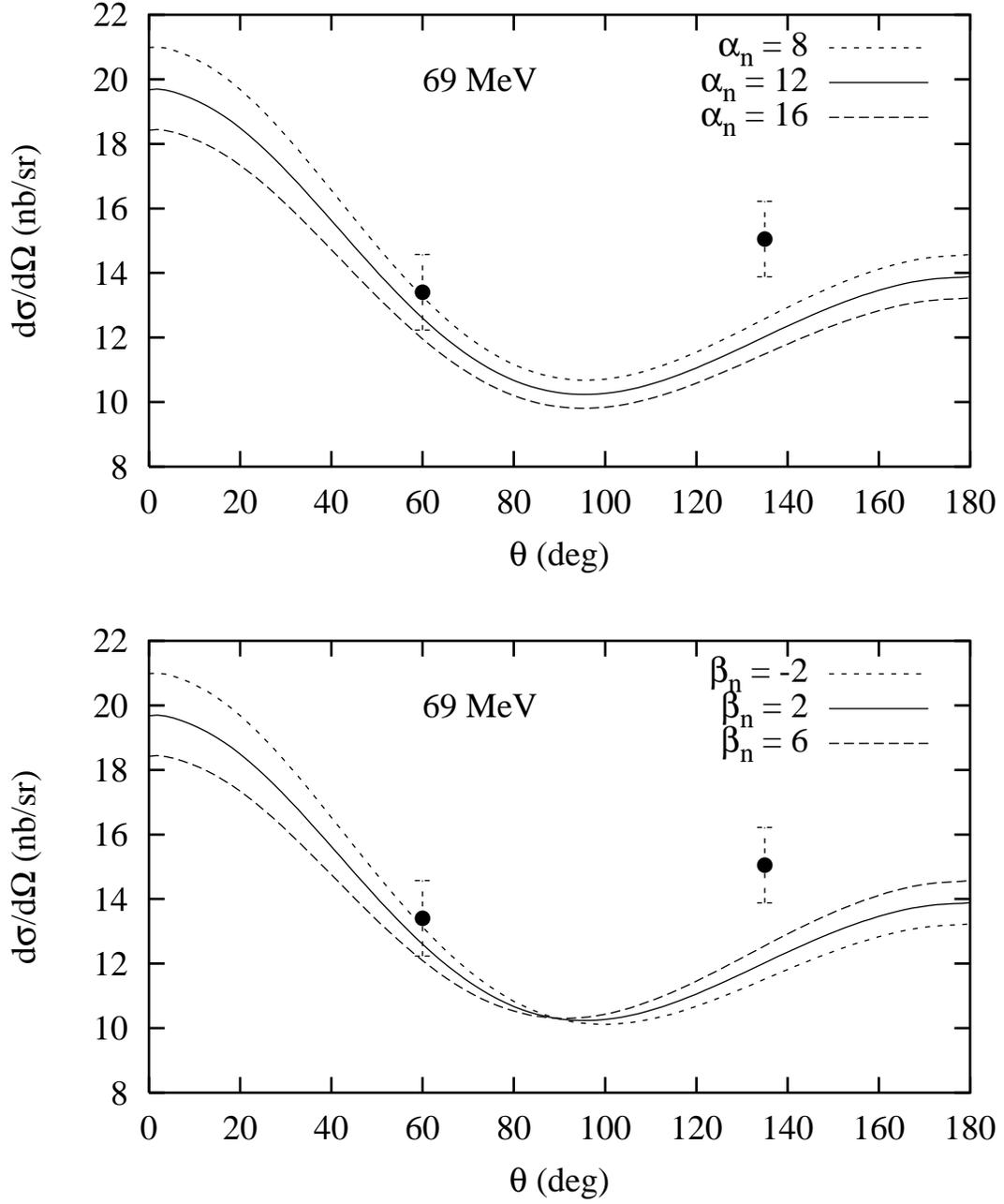

\centering

\caption{ Effect of varying $\alpha_n$
  and $\beta_n$ separately.  $\beta_n = 2.0$ in the top graph and
$\alpha_n = 12.0$ in the bottom.  \label{fig:e2-1} }
\end{figure}

\begin{figure}
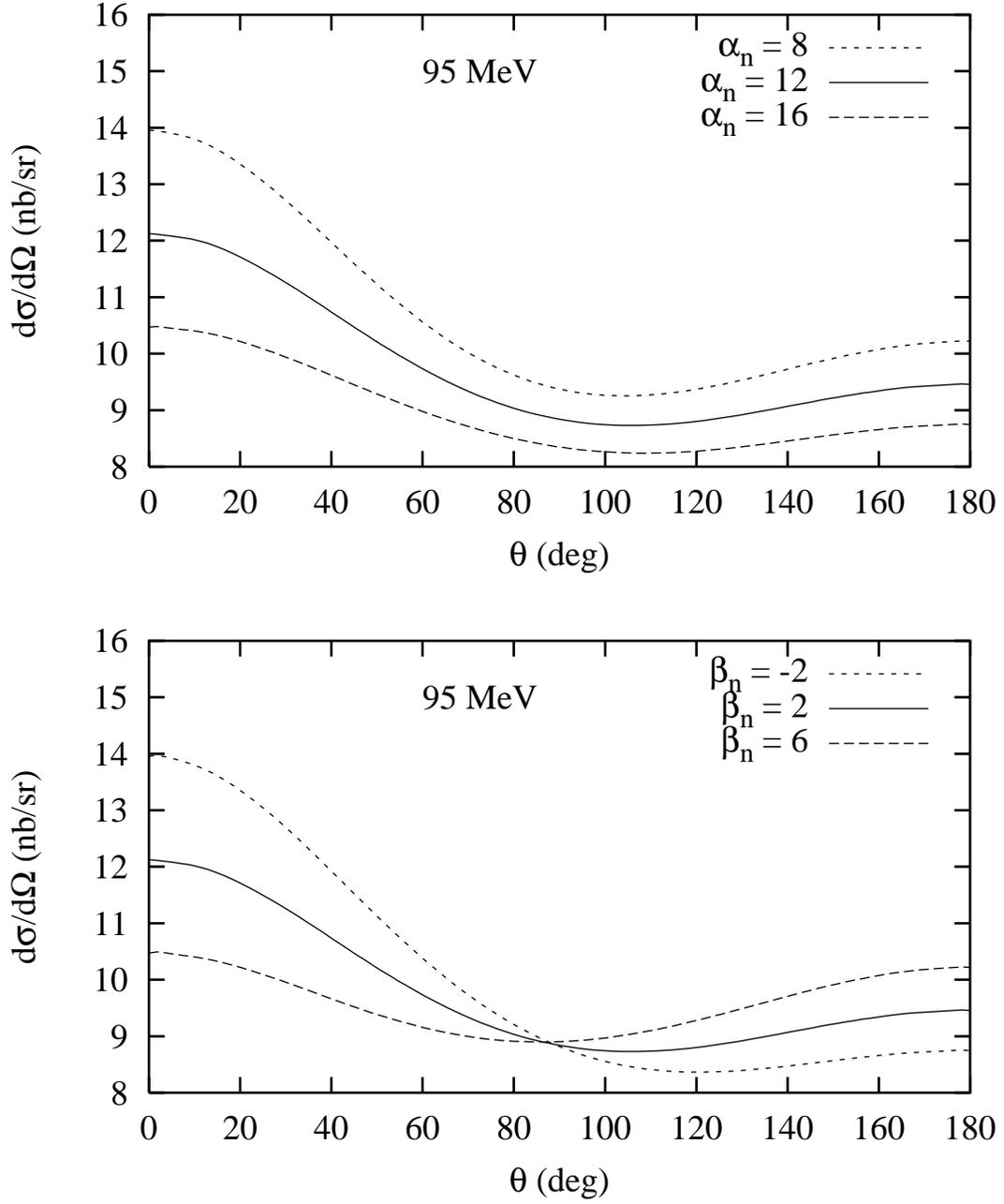

\centering
\caption{ Effect of varying $\alpha_n$
  and $\beta_n$ separately at 95 MeV.  $\alpha_n,\beta_n $ are as in
  Fig.~7 }
\end{figure}
\begin{figure}
\centering
\caption{Effect of varying both $\alpha_n$ and $\beta_n$ . \label{fig:e2-2} }
\end{figure}
\begin{figure}
\centering
\caption{Effect of varying both $\alpha_p$ and $\beta_p$.   \label{fig:e2-3} }
\end{figure}
\begin{figure}
\centering
\caption{Sensitivity to values of $\alpha,\beta$ at 95 MeV} 
\end{figure}
\begin{figure}
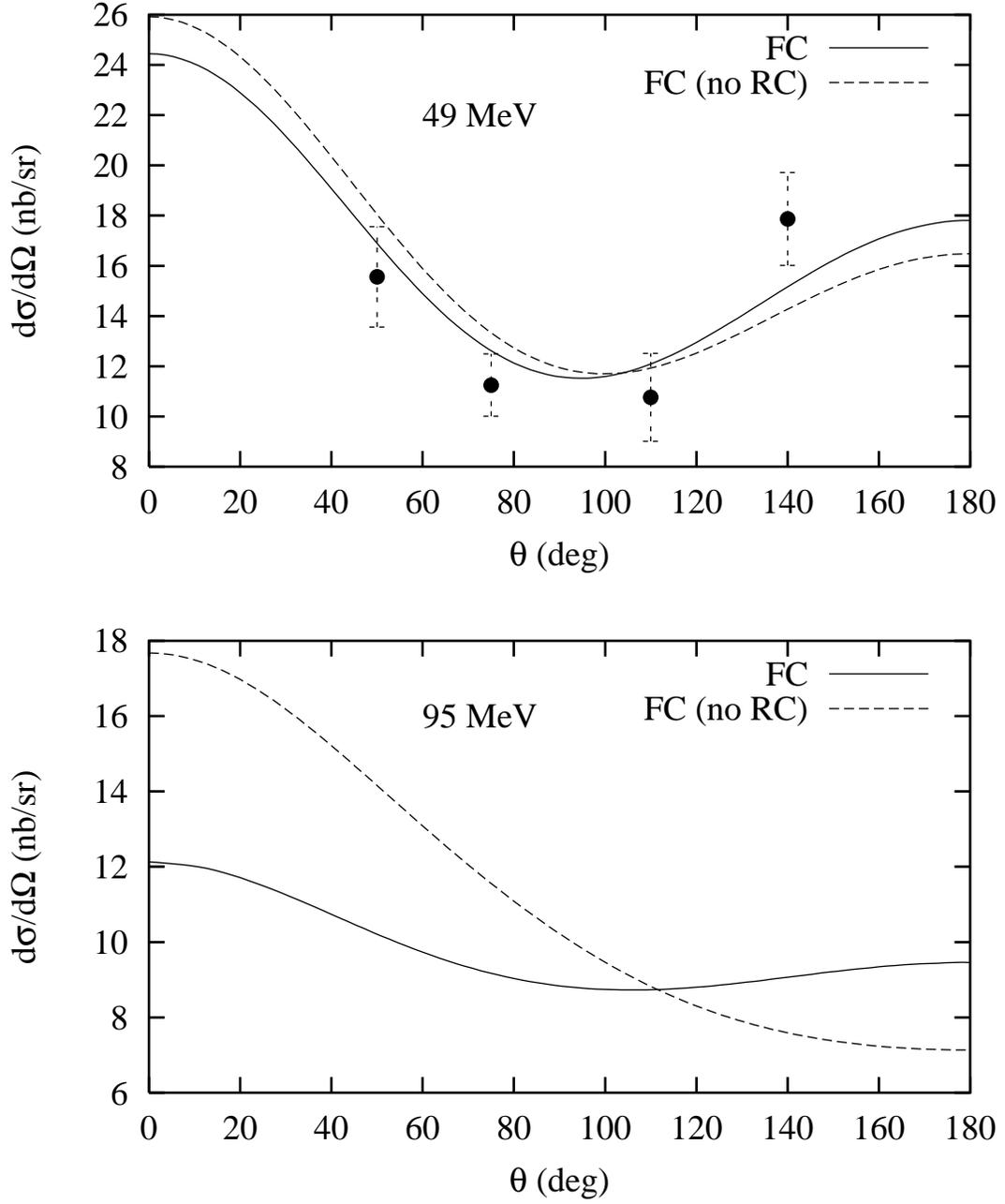

\centering
\caption{ \label{fig:e3-12}
  Effect of neglecting relativistic  terms RC
  on the differential cross-section. The full calculation is labelled FC. }
\end{figure}
\begin{figure}
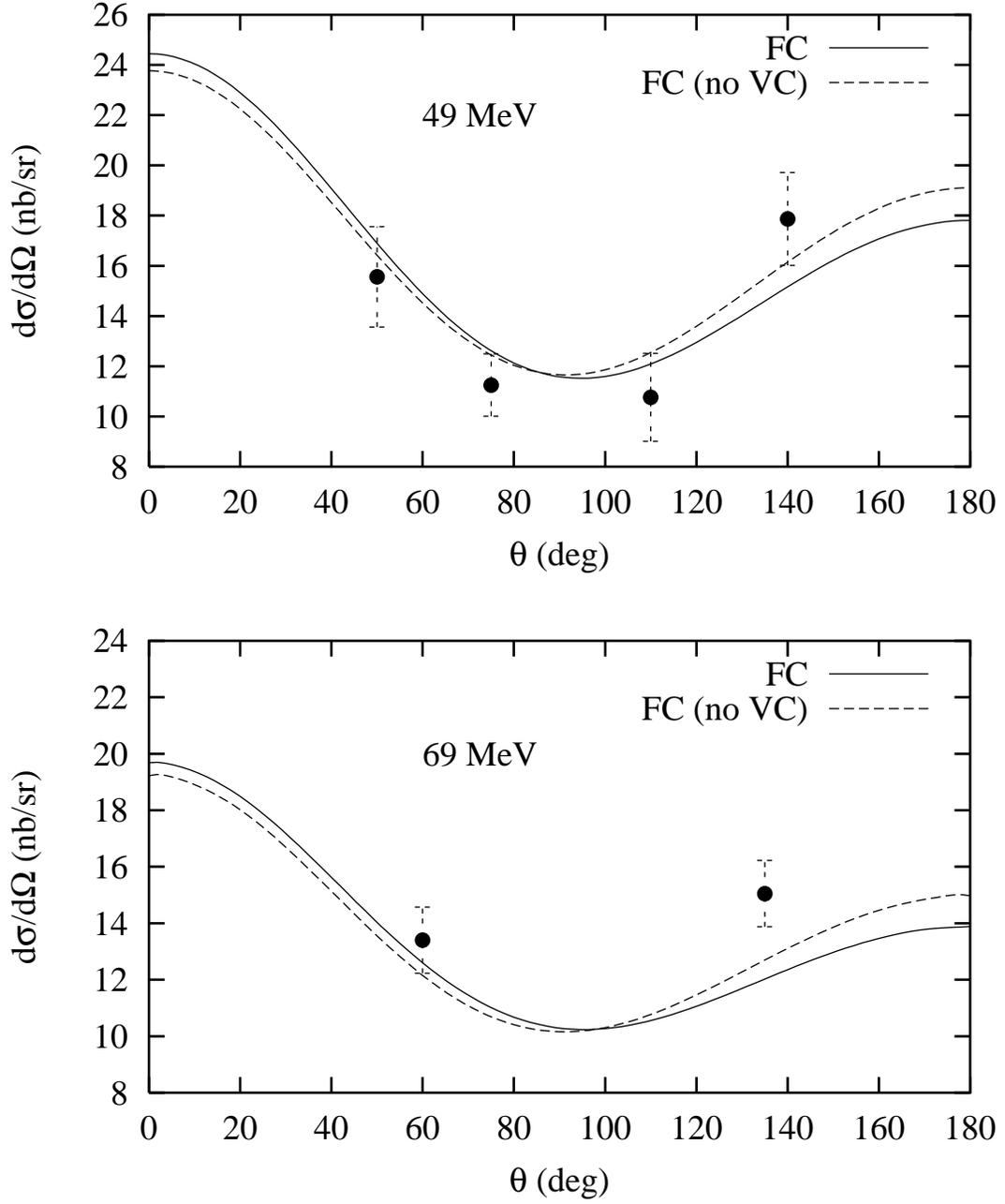

\centering
\caption{ \label{fig:e3-13}
  Effect of neglecting vertex corrections VC
  on the differential cross-section }
\end{figure}
\begin{figure}
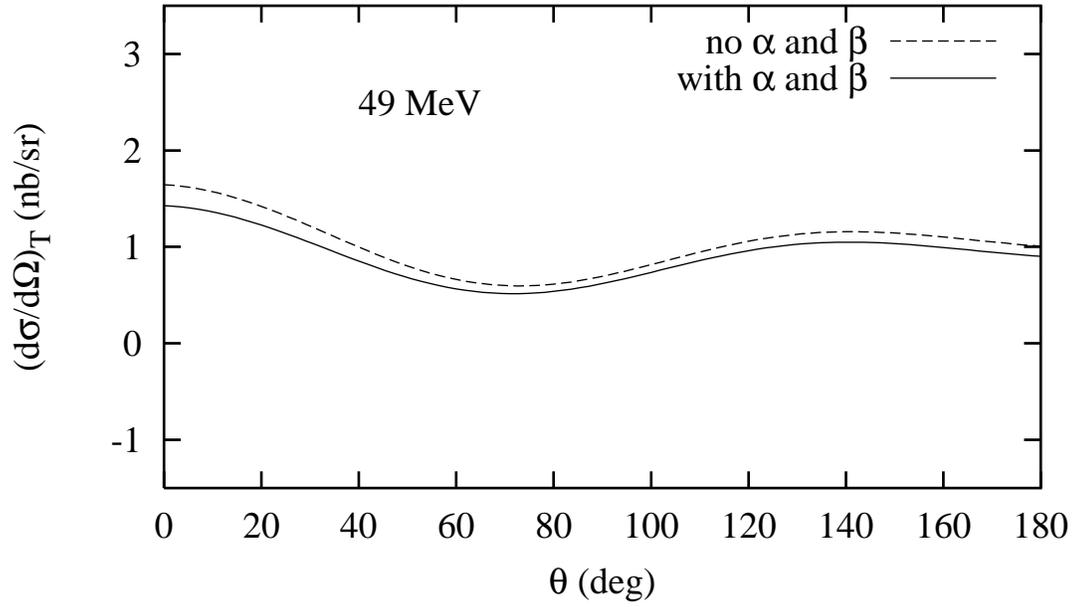

\centering
\caption[Tensor-polarized deuteron Compton scattering cross-section, with and without polarizabilities]{Tensor-polarized deuteron 
Compton scattering cross-section, with and without polarizabilities.  The values of $\alpha_n$ = 12.0, $\beta_n$ = 2.0, $\alpha_p$ = 10.9,
and $\beta_p$ = 3.3 are used in the solid curve.
All other interactions are included.         \label{fig:e4-1} }
\end{figure}
\begin{figure}
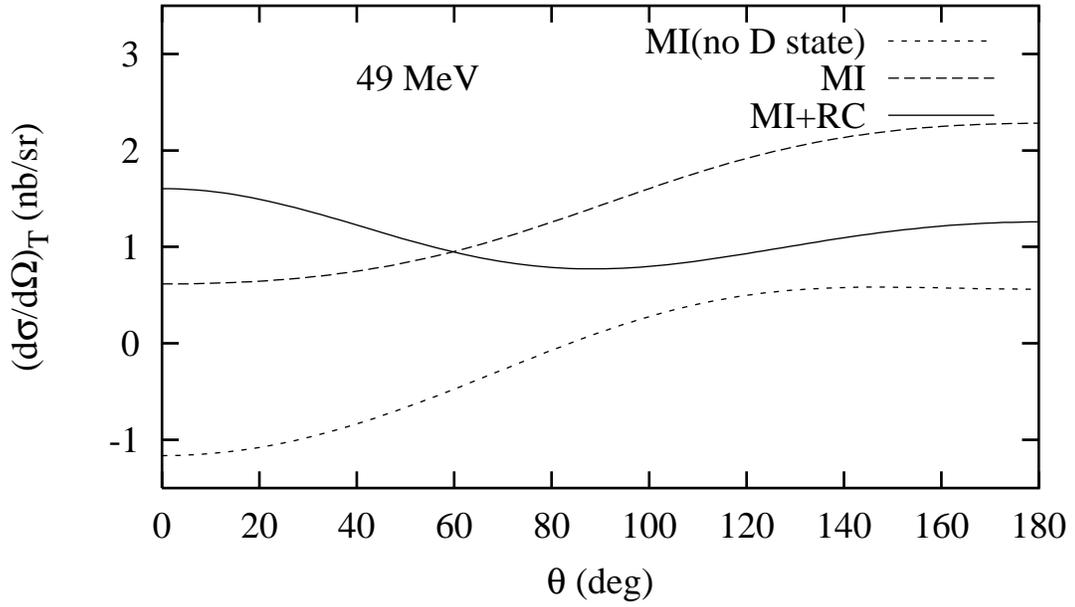

\centering
\caption[Tensor-polarized deuteron Compton scattering cross-section,
including only magnetic interactions]{ Tensor-polarized deuteron 
Compton scattering cross-section, including
only magnetic interactions(MI).  The dotted curve contains
all of the dispersive terms, 
 but only the contributions from the deuteron $S$-state are included.
These contributions
are added in the dashed curve.   Relativistic corrections (RC) are added
in the solid curve. \label{fig:e4-2} }
\end{figure}
\appendix
\section{Leading-Order Dispersive Terms \label{app:main} }
\noindent The remaining terms will be calculated using
second-order perturbation theory:
\begin{eqnarray}
{\cal T}_{fi}^{disp} 
& = & \sum_{C, \vec{P}_C} \left\{ \frac{ \mx{d_f, \vec{P}_f ,{\gamma}_f}{ 
H^{\mathrm{int}} }{C, \vec{P}_C} 
\mx{C, \vec{P}_C}{H^{\mathrm{int}}}{d_i,
  \vec{P}_i, {\gamma}_i} }{\hbar{\omega}_i+E_{d_i}-E_C-
P_C^2/2m_d +i\varepsilon} 
+ \right.  \label{eq:mainterm1}\\
& & \left. \frac{ \mx{d_f, \vec{P}_f, {\gamma}_f}
    {H^{\mathrm{int}}}{C, \vec{P}_C, {\gamma}_i, {\gamma}_f }
\mx{C, \vec{P}_C, {\gamma}_i, {\gamma}_f }{H^{\mathrm{int}}}{d_i,
  \vec{P}_i, {\gamma}_i} } 
{-\hbar{\omega}_f+E_{d_i}-E_C-P_C^2/2m_d+i\varepsilon} \right\} \nonumber .
\end{eqnarray}
\noindent Here $C$ denotes the internal
quantum numbers of the intermediate $np$ state, 
and $\sum_C$
is shorthand for all sums and integrals
which
are needed to
describe this complete set of states except for the center-of-mass states.
These  are 
written separately as $\vec{P}_C$.
At this point, it is most convenient to 
use $H^{\mathrm{int}} = -\int \vec{J}(\vec{\dum})\cdot
\vec{A}(\vec{\dum}) d^3\dum$,
and then expand $\vec{A}$
into multipoles according to equation~(\ref{eq:d4-2}).  There are
3 terms in this expansion; therefore, equation~(\ref{eq:mainterm1})
contains a total of 18 terms. 
The largest contributions at low energies (which we 
denote as  $\matrixm{a}$) arise from the gradient operator in 
Eq.~(\ref{eq:d4-2}).
These are the leading-order contributions; the other
terms are of order $\omega/m_N$ smaller.  This gradient term
includes the case where there is an $E1$ interaction
at both $\gamma N$ vertices.

With this in mind, 
we define $\Phi_i(\vec{r})$ and ${\Phi}_f(\vec{r})$ by 
\begin{eqnarray}
\Phi_i(\vec{r}) & \equiv & -\frac{1}{\sqrt{V}} \sqrt{\frac{2\pi\hbar}{\winit} } \sum_{L=1}^{\infty} \sum_{M=-L}^{L} \wignerdi 
\frac{i^{L+1}}{\winit} \sqrt{ \frac{2\pi(2L+1)}{L(L+1)} }  \times
\label{eq:mainterm10} \\
& & \ \ \ \ \ \ \left( 1+r\frac{d}{dr} \right) j_{\sss L}(\winit r) Y_{LM}(\hat{r}) \, \nonumber \\
\Phi_f(\vec{r}) & \equiv & \frac{1}{\sqrt{V}} \sqrt{\frac{2\pi\hbar}{\wfin} } \sum_{L'=1}^{\infty} \sum_{M'=-L'}^{L'}
 (-1)^{L'-\mufin} \wignerdf
\frac{i^{L'+1}}{\wfin} \times \nonumber \\
& & \ \ \ \ \ \  \sqrt{ \frac{2\pi(2L'+1)}{L'(L'+1)} }
\left( 1+r\frac{d}{dr} \right) j_{\sss L'}(\wfin r) Y_{L'M'}(\hat{r}),
\label{eq:mainterm11}
\end{eqnarray}
\noindent which means that 
\begin{eqnarray}
\frac{1}{\sqrt{V}} \sqrt{\frac{2\pi\hbar}{\winit} } \eps 
e^{i\kinit\cdot\vec{r}} \label{eq:mainterm2} & = &
\vec{\nabla}_r\Phi_i(\vec{r}) + \cdots, \\
\frac{1}{\sqrt{V}} \sqrt{\frac{2\pi\hbar}{\wfin} } \epspr 
e^{-i\kfin\cdot\vec{r}} \label{eq:mainterm3}
& = & \vec{\nabla}_r\Phi_f(\vec{r}) + \cdots.
\end{eqnarray}

After a lengthy calculation\cite{thesis}
one  finds
\begin{equation}
\matrixm{a}  = 
\matrixm{a1} + \matrixm{a2} + \matrixm{a3} + \matrixm{a4},  
\end{equation}
\noindent where 
\begin{eqnarray}
\lefteqn{ \matrixm{a1}   =   \frac{-2\pi e^2}{(\hbar\winit)(\hbar\wfin)} 
\left[ \frac{(\hbar\winit)^2}{2m_d} - \hbar\winit \right]^2
\sum_{l=0,2} \sum_{l'=0,2} \sum_{L,L'=1}^{\infty} \sum_{S_C=0,1} 
\sum_{J_C=|1-L|}^{1+L} \sum_{L_C=|J_C-S_C|}^{J_C+S_C}\times }\nonumber\\
& & \ \ \ \ i^{L+L'}(-1)^{L'-\mufin-M_f+J_C-M_i-\muinit} \wignerdfnom
\times\nonumber \\
& & \ \ \ \ \sqrt{\frac{(2L+1)(2L'+1)}{LL'(L+1)(L'+1)}}
\threej{1}{L'}{J_C}{-M_f}{M_f-M_i-\muinit}{\muinit+M_i} \times \nonumber \\
& & \ \ \ \ \threej{J_C}{L}{1}{-\muinit-M_i}{\muinit}{M_i} \mxredbc{Y_{L'}}
\mxredca{ Y_{L} } \times\nonumber\\ 
& & \ \ \ \ \int dr\  r\  u_{\sss l}(r) \chi_f^{l'\hat{C}}(r) 
\psi_{\sss L}(\frac{\winit r}{2}).  
\end{eqnarray}
We have used the definitions
$E_0 \equiv \hbar\winit -  \frac{(\hbar\winit)^2}{2m_d} - E_b$, 
and $ 
\psi_{\sss L}(x) \equiv j_{\sss L}(x) + x \frac{d}{dx} j_{\sss L}(x).
$ 
\begin{eqnarray}
\matrixm{a2} & = & \frac{-2\pi e^2}{(\hbar\winit)(\hbar\wfin)} 
\left[ \hbar\wfin + \frac{ (\hbar\wfin)^2 }{2m_d} \right]^2 
\sum_{l=0,2} \sum_{l'=0,2} \sum_{L,L'=1}^{\infty}  \sum_{S_C=0,1}
\sum_{J_C=|1-L|}^{1+L} \sum_{L_C=|J_C-S_C|}^{J_C+S_C}\times \nonumber\\ 
& & i^{L+L'}(-1)^{L'-\mufin+J_C-\muinit} 
\wignerdfnom  \sqrt{\frac{(2L+1)(2L'+1)}{LL'(L+1)(L'+1)}} \times\nonumber\\
& & \threej{1}{L}{J_C}{-M_f}{\muinit}{-\muinit+M_f} \threej{J_C}{L'}{1}{\muinit-M_f}{M_f-M_i-\muinit}{M_i} 
\times \nonumber \\
& & \mxredbc{Y_{L}} \mxredca{ Y_{L'} } \times\nonumber\\
& & \int  dr\  r u_{\sss l'}(r) \chi_f^{l\hat{C}}(r) 
 \psi_{\sss L}(\frac{\winit r}{2}),
\end{eqnarray}
\noindent
where $E_0' \equiv -\hbar\wfin -  \frac{ (\hbar\wfin)^2 }{2m_d} - E_b$.
\begin{eqnarray}
\lefteqn{ \matrixm{a3}
 =  -\frac{e^2 \sqrt{\pi}}{(\hbar\wfin)(\hbar\winit)} 
\left[ \hbar\wfin  + \frac{ (\hbar\wfin)^2 }{2m_d} - \hbar\winit +  \frac{ (\hbar\winit)^2 }{2m_d} \right] 
\sum_{l=0,2} \sum_{l'=0,2} \sum_{\tilde{L}=|L-L'|}^{L+L'} \sum_{L=1}^{\infty} \sum_{L'=1}^{\infty} \times} \nonumber\\
& & \wignerdfnom (2L+1)(2L'+1) \sqrt{\frac{2\tilde{L}+1}{LL'(L+1)(L'+1)}} \times \nonumber \\
& &  \threej{L}{L'}{\tilde{L}}{0}{0}{0} 
\threej{L}{L'}{\tilde{L}}{\muinit}{M_f-M_i-\muinit}{M_i-M_f} \threej{1}{\tilde{L}}{1}{-M_f}{M_f-M_i}{M_i} 
\times \nonumber \\
& &  i^{L+L'} (-1)^{L'-\mufin -M_i}  \mxredba{Y_{\tilde{L} } }
\int dr u_{\sss l}(r) u_{\sss l'}(r) \psi_{\sss L'}(\frac{\wfin r}{2}) \psi_{\sss L}(\frac{\winit r}{2}).
\end{eqnarray}
\noindent The final term involves the double commutator of
the Hamiltonian. The
 kinetic energy contribution is given here  and potential energy below.
\begin{eqnarray}
\matrixm{a4, \mathrm{KE}} & = & \frac{e^2\sqrt{\pi}}{m_p \wfin\winit} 
\sum_{l=0,2} \sum_{l'=0,2} \sum_{\tilde{L}=|L-L'|}^{L+L'} \sum_{L=1}^{\infty} \sum_{L'=1}^{\infty} i^{L+L'} (-1)^{L'-\mufin - M_i}
\times\nonumber\\
& &  \ \ \ \ (2L+1)(2L'+1 ) \sqrt{ \frac{2\tilde{L}+1}{LL'(L+1)(L'+1)} }
\threej{L}{L'}{\tilde{L}}{0}{0}{0} \nonumber \times \\
& & \ \ \ \ \threej{L}{L'}{\tilde{L}}{\muinit}{M_f-M_i-\muinit}{M_i-M_f} 
\threej{1}{\tilde{L}}{1}{-M_f}{M_f-M_i}{M_i} \nonumber \times \\
& & \ \ \ \ \wignerdfnom \mxredba{  Y_{\tilde{L}} }  \times \nonumber \\
& & \ \ \ \ \int dr\ u_{\sss l}(r) u_{\sss l'}(r) 
\left\{ 2 \left[ \frac{d}{dr}\psi_{\sss L}(\frac{\winit r}{2}) \right]
\left[ \frac{d}{dr}\psi_{\sss L'}(\frac{\wfin r}{2}) \right] + \nonumber \right. \\
& & \ \ \ \ \left. \frac{ L(L+1) + L'(L'+1) - \tilde{L}(\tilde{L}+1) }{r^2} \psi_{\sss L}(\frac{\winit r}{2})
\psi_{\sss L'}(\frac{\wfin r}{2}) \right\}.
\end{eqnarray}
The other dispersive terms are listed in \cite{thesis}. 

\section{Recoil Corrections \label{app:recoil} }
\noindent We will now return to the recoil
corrections to the
transition matrices which arise from the $\vec{P}$ operators in
Eqs.~(\ref{eq:d4-1} )
and (\ref{eq:d4-3}) 
The main contributions from these terms are 
\begin{eqnarray}
\lefteqn{ \matrixt{\mathrm{cm, uncr}}  \equiv } \\
& &  \sum_{C, \vec{P}_C} 
\frac{ \mx{d_f, \vec{P}_f }{ \frac{ie}{m_p} \vec{\nabla}\Phi_f(\frac{\vec{r}}{2})
e^{-i\vec{k}_f\cdot\vec{R} }\cdot\vec{P} }{C, \vec{P}_C} \mx{C, \vec{P}_C}{\left[ H^{np}, \hat{\Phi}_i \right]
e^{i\vec{k}_i\cdot\vec{R}}}{d_i,\vec{P}_i} }{
\hbar\winit + E_{d_i} -E_C - P_C^2/2m_d + i\varepsilon }, \nonumber\\ 
\lefteqn{ \matrixt{\mathrm{cm,cr} }   \equiv  } \\
& & \sum_{C, \vec{P}_C} 
\frac{ \mx{d_f, \vec{P}_f }{ \frac{ie}{m_p} \vec{\nabla}\Phi_i(\frac{\vec{r}}{2})
e^{i\vec{k}_i\cdot\vec{R} }\cdot\vec{P} }{C, \vec{P}_C} \mx{C, \vec{P}_C}{\left[ H^{np}, \hat{\Phi}_f \right]
e^{-i\vec{k}_f\cdot\vec{R}}}{d_i,\vec{P}_i} }{
-\hbar\wfin  + E_{d_i} -E_C - P_C^2/2m_d + i\varepsilon }. \nonumber
\end{eqnarray}
\noindent Since we are working in the lab frame, $\vec{P} \ket{\vec{P}_i} = 0.$
We  integrate the center-of-mass variables, use $\vec{A} = \vec{\nabla}\Phi$,
and evaluate the commutator so that
 $\matrixt{\mathrm{cm, uncr}}$ becomes:,
\begin{eqnarray}
\matrixt{\mathrm{cm, uncr}} & = &   \sqrt{ \frac{2\pi\hbar}{V\wfin} }
\frac{ \delta(\vec{P}_i + \vec{k}_i - \vec{P}_f - \vec{k}_f)  }{V} \left[
\matrixt{\mathrm{cm, uncr}1} + \matrixt{\mathrm{cm, uncr}2} \right],
\end{eqnarray}
\noindent where
\begin{eqnarray}
\matrixt{\mathrm{cm, uncr}1} & = & \sum_C  \frac{ \mx{d_f}{ \frac{ie\hbar}{m_p} (\vec{k}_i\cdot\epspr )
e^{-\frac{i}{2}\vec{k}_f\cdot\vec{r} } }{C}
\mx{C}{ \left[ \hbar\winit - \frac{(\hbar\winit)^2}{2m_d} \right] \hat{\Phi}_i }{d_i} }{ \hbar\winit -
\frac{(\hbar\winit)^2}{2m_d}+ E_{d_i} -E_C + i\varepsilon },\nonumber\\
\matrixt{\mathrm{cm, uncr}2} & = & -\mx{d_f}{ \frac{ie\hbar}{m_p} (\vec{k}_i\cdot\epspr)
e^{-\frac{i}{2}\vec{k}_f\cdot\vec{r} } \hat{\Phi}_i }{d_i}.
\end{eqnarray}
\noindent Starting with $\matrixt{\mathrm{cm, uncr}1}$, we do some algebra
to obtain:
\begin{eqnarray}
\lefteqn{\matrixm{\mathrm{cm, uncr}1} = -\sum_{l=0,2} \sum_{l'=0,2} \sum_{S_C = 0,1} \sum_{J_C = |1-L|}^{1+L} \sum_{L_C=|J_C-S_C|}^{J_C+S_C}
 \sum_{L=1}^{\infty} \sum_{L'=0}^{\infty} \times}\\
& & (\hat{k}_i\cdot\epspr) \frac{4\pi e^2}{m_p} \left[ \hbar\winit- \frac{(\hbar\winit)^2}{2m_d} \right] (-1)^{J_C-M_f-M_i-\muinit}
\times\nonumber \\
& & i^{L-L'} Y_{L', M_f-M_i-\muinit}^{\ast}(\hat{k}_f) 
\sqrt{ \frac{2\pi(2L+1)}{L(L+1)} }\times\nonumber\\
& & \mxredbc{Y_{L'}} \mxredca{Y_L} \times\nonumber\\
& & \threej{1}{L'}{J_C}{-M_f}{M_f-M_i-\muinit}{M_i+\muinit} \threej{J_C}{L}{1}{-M_i-\muinit}{\muinit}{M_i} \times\nonumber\\
& & \int_{0}^{\infty} r\ dr  \chi_f^{l'\hat{C}}(r) 
\psi_{\sss L}(\frac{\winit r}{2})  u_{\sss l}(r).\nonumber
\end{eqnarray}
\noindent We  list the results for the other
$\matrixm{\mathrm{cm}}$ terms:
\begin{eqnarray}
\matrixm{\mathrm{cm, uncr}2} & = &  \sum_{l=0,2} \sum_{l'=0,2} \sum_{\tilde{L}=|L-L'|}^{L+L' }\sum_{L=1}^{\infty} \sum_{L'=0}^{\infty} 
(\hat{k}_i\cdot\epspr) \frac{4\pi e^2}{m_p} (-1)^{M_i} \times\\
& & i^{L-L'} Y_{L', M_f-M_i-\muinit}^{\ast}(\hat{k}_f)
(2L+1) \sqrt{ \frac{(2L'+1)(2\tilde{L}+1)}{2L(L+1)}} \times\nonumber\\
& & \mxredba{Y_{\tilde{L}}} \threej{L}{L'}{\tilde{L}}{0}{0}{0} \threej{L}{L'}{\tilde{L}}{\muinit}{M_f-M_i-\muinit}{M_i-M_f}
\times\nonumber\\
& & \threej{1}{\tilde{L}}{1}{-M_f}{M_f-M_i}{M_i} \int_0^\infty
dr\  u_{\sss l'}(r) j_{\sss L'}(\frac{\wfin r'}{2})
\psi_{\sss L}(\frac{\winit r}{2})  u_{\sss l}(r),\nonumber\\
\matrixm{\mathrm{cm, cr}1} & = &  \sum_{l=0,2} \sum_{l'=0,2} \sum_{S_C = 0,1} \sum_{J_C = |1-L|}^{1+L} \sum_{L_C=|J_C-S_C|}^{J_C+S_C}
\sum_{L'=1}^{\infty} \sum_{L=0}^{\infty}  \times\\
& &  (\hat{k}_f\cdot\eps) \frac{4\pi e^2}{m_p} \left[ \hbar\wfin+\frac{(\hbar\wfin)^2}{2m_d} \right] (-1)^{L'-\mufin-J_C}
\times\nonumber\\
& & i^{L+L'} 
\sqrt{ \frac{(2L+1)(2L'+1)}{2L'(L'+1)}}\wignerdf\times\nonumber\\
& & \mxredbc{Y_{L}} \mxredca{Y_{L'}} \times\nonumber\\
& & \threej{1}{L}{J_C}{-M_f}{0}{M_f}
\threej{J_C}{L'}{1}{-M_f}{M_f-M_i}{M_i}  \times\nonumber\\
& & \int_{0}^{\infty} r\ dr
\chi_f^{l\hat{C}}(r)
j_{\sss L}(\frac{\winit r}{2})  u_{\sss l'}(r),\nonumber\\
\matrixm{\mathrm{cm, cr}2} & = & \sum_{\tilde{L}=|L-L'|}^{L+L' } \sum_{l=0,2} \sum_{l'=0,2} \sum_{L=0}^{\infty} \sum_{L'=1}^{\infty} 
(\hat{k}_f\cdot\eps) \frac{2 e^2}{m_p}  (-1)^{L'-\mufin-M_i} \times\\
& & i^{L+L'} (2L'+1) (2L+1)
\sqrt{ \frac{\pi(2\tilde{L}+1)}{2L'(L'+1)} } \wignerdf 
\times\nonumber\\
& & \mxredba{Y_{\tilde{L}}} \threej{L}{L'}{\tilde{L}}{0}{0}{0} \threej{L}{L'}{\tilde{L}}{0}{M_f-M_i}{M_i-M_f}
\times\nonumber\\
& & \threej{1}{\tilde{L}}{1}{-M_f}{M_f-M_i}{M_i} \int_{0}^{\infty}  dr\  u_{\sss l'}(r)  j_{\sss L}(\frac{\winit r}{2})
\psi_{\sss L'}(\frac{\wfin r}{2})  u_{\sss l}(r). \nonumber
\end{eqnarray}

\section{Vertex Corrections} \label{app:pion} 

We calculate terms,  such as the one shown            
in Fig.~1f,, 
in which a $(\gamma,\pi) $ reaction on one
nucleon is followed by
the absorption of the pion on the second. This  is followed (or preceeded) by
a photon emission from the second nucleon.
These are vertex corrections to the ``ordinary'' second-order diagrams.

The  result is
\begin{eqnarray}
\lefteqn{ \matrixm{\pi \mathrm{vc} 1 }  =  -\sum_{l=0,2} \sum_{l'=0,2} \sum_{\tilde{J}=|1-J_C|}^{1+J_C} 
\sum_{J_C=|1-L''|}^{1+L''} \sum_{L''=|L-L'|}^{L+L'}
\sum_{L=0}^{\infty} \sum_{L'=1}^{\infty} \sum_{i=-1,0,1} \sum_{j=-1,0,1} \times}\\
& & \frac{72\pi e^2 f^2 \hbar^2 \mufin\wfin}{m_p m_{\pi}^2}
i^{L+L'} (-1)^{L'-\mufin-M_f-M_i+j+J_C-L''} \left[ \mu_p + (-1)^{L'} \mu_n \right]  \times\nonumber\\
& & \wignerdf (\eps)_{-i} (2\tilde{J}+1)(2L+1) \sqrt{ \frac{(L'+1)(2L''+1)(2J_C+1)}{4\pi} }
\times\nonumber\\
& & \threej{1}{L'}{J_C}{-M_f}{M_f-M_i-i}{M_i+i} \threej{1}{1}{1}{j}{i}{-i-j}  \threej{1}{L}{L''}{0}{0}{0} \times\nonumber\\
& & \threej{1}{L}{L''}{-j}{0}{j} \threej{1}{L''}{\tilde{J}}{i+j}{-j}{-i} \threej{J_C}{\tilde{J}}{1}{-M_i-i}{i}{M_i} \times\nonumber\\
& & \ninej{0}{J_C}{J_C}{1}{l}{1}{1}{L''}{\tilde{J}} \mxred{l' 1 1}{  [ Y_{L'-1} \otimes t ]_{L'} }{J_C 0 J_C}
\mxred{J_C}{ Y_{L''} }{l} \times\nonumber\\
& & \int_0^{\infty} \int_0^{\infty} r\ dr\ r'\ dr'\ j_{\sss L'-1}(\frac{\wfin r'}{2}) u_{\sss l'}(r') \green j_{\sss L}(\frac{\winit r}{2})
\frac{d}{dr} \frac{e^{-m_{\pi}r/\hbar} }{r} u_{\sss l}(r),\nonumber
\end{eqnarray}
where the Green's function $\green$ is defined by
\begin{equation}
  \green\equiv\langle r'\mid{1\over E_0-H^{np}_{\hat C}+i\epsilon}
  \mid r\rangle.
  \end{equation}
The other correction terms are:
\begin{eqnarray}
\lefteqn{ \matrixm{\pi \mathrm{vc} 2 }  =  -\sum_{l=0,2} \sum_{l'=0,2} \sum_{\tilde{J}=|1-J_C|}^{1+J_C}
\sum_{J_C=|1-L''|}^{1+L''} \sum_{L''=|L-L'|}^{L+L'}
\sum_{L=0}^{\infty} \sum_{L'=1}^{\infty} \sum_{i=-1,0,1} \sum_{j=-1,0,1} \times}\\
& &  \frac{72\pi e^2 f^2 \hbar^2 \muinit\winit}{m_p m_{\pi}^2}
i^{L+L'} (-1)^{L'+i+j-L''-M_f+J_C-M_i-\muinit} \left[ \mu_p + (-1)^{L} \mu_n \right]\times\nonumber\\
& &   Y_{L'M'}^{\ast}(\hat{k}_f) (\epspr)_{-i}
(2\tilde{J}+1) \sqrt{ (L+1)(2L'+1)(2L''+1)(2J_C+1) } \times\nonumber\\
& & \threej{1}{L'}{L''}{0}{0}{0}\threej{1}{L'}{L''}{-j}{M_f-M_i-\muinit-i}{M_i+\muinit+i+j-M_f} \times\nonumber\\
& & \threej{1}{1}{1}{j}{i}{-i-j}\threej{1}{L''}{\tilde{J}}{i+j}{M_f-M_i-\muinit-i-j}{\muinit+M_i-M_f} \times\nonumber\\
& & \threej{J_C}{L}{1}{-M_i-\muinit}{\muinit}{M_i}\threej{1}{\tilde{J}}{J_C}{-M_f}{M_f-M_i-\muinit}{M_i+\muinit} \times\nonumber\\
& & \ninej{1}{l'}{1}{0}{J_C}{J_C}{1}{L''}{\tilde{J}} \mxred{J_C 0 J_C}{  [ Y_{L-1} \otimes t ]_{L} }{l 1 1}
\mxred{l'}{ Y_{L''} }{J_C} \times\nonumber\\
& & \int_0^{\infty} \int_0^{\infty} r\ dr\ r'\ dr'\ j_{\sss L-1}(\frac{\winit r}{2}) u_{\sss l}(r) \green j_{\sss L'}(\frac{\wfin r'}{2})
\frac{d}{dr'} \frac{e^{-m_{\pi}r'/\hbar} }{r'} u_{\sss l'}(r')\nonumber ,
\end{eqnarray}
\begin{eqnarray} 
\lefteqn{\matrixm{\pi\mathrm{vc} 3 }  =  -\sum_{l=0,2} \sum_{l'=0,2} \sum_{\tilde{J}=|1-J_C|}^{1+J_C}
\sum_{J_C=|1-L''|}^{1+L''} \sum_{L''=|L-L'|}^{L+L'}
\sum_{L=0}^{\infty} \sum_{L'=1}^{\infty} \sum_{i=-1,0,1} \sum_{j=-1,0,1} \times}\\
& & \frac{72\pi e^2 f^2 \hbar^2 \mufin\wfin}{m_p m_{\pi}^2}
i^{L+L'} (-1)^{L'-\mufin+J_C+j+L''}  \left[ \mu_p + (-1)^{L'} \mu_n \right]\times\nonumber\\
& & \wignerdf (\eps)_{-i}
(2\tilde{J}+1)(2L+1) \sqrt{ \frac{(L'+1)(2L''+1)(2J_C+1)}{4\pi} } \times\nonumber\\
& & \threej{J_C}{L'}{1}{i-M_f}{M_f-M_i-i}{M_i} \threej{1}{1}{1}{j}{i}{-i-j}  \threej{1}{L}{L''}{0}{0}{0} \times\nonumber\\
& & \threej{1}{L}{L''}{-j}{0}{j} \threej{1}{L''}{\tilde{J}}{i+j}{-j}{-i}
\threej{1}{\tilde{J}}{J_C}{-M_f}{i}{M_f-i} \times\nonumber\\
& & \ninej{1}{l'}{1}{0}{J_C}{J_C}{1}{L''}{\tilde{J}} \mxred{J_C 0 J_C}{  [ Y_{L'-1} \otimes t ]_{L'} }{l 1 1}
\mxred{l'}{ Y_{L''} }{J_C} \times\nonumber\\
& & \int_0^{\infty} \int_0^{\infty} r\ dr\ r'\ dr'\ j_{\sss L'-1}(\frac{\wfin r'}{2}) u_{\sss l}(r') \greenpr j_{\sss L}(\frac{\winit r}{2})
\frac{d}{dr} \frac{e^{-m_{\pi}r/\hbar} }{r} u_{\sss l'}(r) \nonumber, 
\end{eqnarray}
\begin{eqnarray}
\lefteqn{ \matrixm{\pi \mathrm{vc} 4 }  =  -\sum_{l=0,2} \sum_{l'=0,2} \sum_{\tilde{J}=|1-J_C|}^{1+J_C}
\sum_{J_C=|1-L''|}^{1+L''} \sum_{L''=|L-L'|}^{L+L'}
\sum_{L=0}^{\infty} \sum_{L'=1}^{\infty} \sum_{i=-1,0,1} \sum_{j=-1,0,1} \times}\\
& & \frac{72\pi e^2 f^2 \hbar^2 \muinit\winit}{m_p m_{\pi}^2}
i^{L+L'} (-1)^{L'+i+j-L''+J_C-\muinit} \left[ \mu_p + (-1)^{L} \mu_n \right]\times\nonumber\\
& &   Y_{L'M'}^{\ast}(\hat{k}_f) (\epspr)_{-i}
(2\tilde{J}+1) \sqrt{ (L+1)(2L'+1)(2L''+1)(2J_C+1) } \times\nonumber\\
& & \threej{1}{L'}{L''}{0}{0}{0}\threej{1}{L'}{L''}{-j}{M_f-M_i-\muinit-i}{M_i+\muinit+i+j-M_f} \times\nonumber\\
& & \threej{1}{1}{1}{j}{i}{-i-j}\threej{1}{L''}{\tilde{J}}{i+j}{M_f-M_i-\muinit-i-j}{\muinit+M_i-M_f}\times\nonumber\\
& & \threej{1}{L}{J_C}{-M_f}{\muinit}{M_f-\muinit}\threej{J_C}{\tilde{J}}{1}{\muinit-M_f}{M_f-M_i-\muinit}{M_i} \times\nonumber\\
& & \ninej{0}{J_C}{J_C}{1}{l'}{1}{1}{L''}{\tilde{J}} \mxred{l'1 1}{  [ Y_{L-1} \otimes t ]_{L} }{J_C 0 J_C}
\mxred{J_C}{ Y_{L''} }{l} \times\nonumber\\
& & \int_0^{\infty} \int_0^{\infty} r\ dr\ r'\ dr'\ j_{\sss L-1}(\frac{\winit r}{2}) u_{\sss l}(r') \greenpr j_{\sss L'}(\frac{\wfin r'}{2})
\frac{d}{dr'} \frac{e^{-m_{\pi}r'/\hbar} }{r'} u_{\sss l}(r').\nonumber
\end{eqnarray}

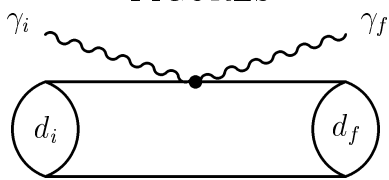
\end{document}